\documentclass[%
	floatfix,
	twocolumn,
	amsfonts,
	aps,
	prd,
	eqsecnum,
	bibtotoc,
	tightenlines,
	twoside,
	showpacs]{revtex4}

\usepackage{psfrag}
\usepackage{graphicx}
\usepackage{amsmath,epsfig,amssymb,amstext,color}
\usepackage{bm}

\newcommand{\N}{\mathcal{N}}
\newcommand{\wn}{\mathfrak{w}}
\newcommand{\mn}{\mathfrak{m}}

\newcommand{\R}{\mathfrak{R}}

\newcommand{\dd}{\mathrm{d}}

\def\RE {I\kern-6pt R    }
\def\Z  {Z\kern-13pt Z   }

\def\bi {\begin{itemize} }

\def\ei {\end{itemize}   }
\def\gtwid{\mathrel{\raise.3ex\hbox{$>$\kern-.75em\lower1ex\hbox{$\sim$}}}}
\def\ltwid{\mathrel{\raise.3ex\hbox{$<$\kern-.75em\lower1ex\hbox{$\sim$}}}}

\begin{document}

\title{{\footnotesize \hfill \rm{MPP-2007-136}}\\[\medskipamount]
Holographic vector mesons from spectral functions\\
at finite baryon or isospin density}

\author{Johanna Erdmenger}
	\email{jke@mppmu.mpg.de}
\author{Matthias Kaminski}
	\email{kaminski@mppmu.mpg.de}
\author{Felix Rust}
	\email{rust@mppmu.mpg.de}
\affiliation{Max-Planck-Institut f\"ur Physik (Werner-Heisenberg-Institut),
         F\"ohringer Ring 6,
         80805 M\"unchen, Germany}

\begin{abstract}
\noindent
We consider gauge/gravity duality with flavor for the finite-temperature field
theory dual of the AdS-Schwarzschild black hole background with embedded
D7-brane probes. In particular, we investigate spectral functions at finite
baryon density in the black hole phase. We determine the resonance frequencies
corresponding to meson-mass peaks as function of the quark mass over temperature
ratio. We find that these frequencies have a minimum for a finite value of the
quark mass. If the quotient of quark mass and temperature is increased further,
the peaks move to larger frequencies. At the same time the peaks narrow, in
agreement with the formation of nearly stable vector meson states which exactly
reproduce the meson mass spectrum found at zero temperature. We also calculate
the diffusion coefficient, which has finite value for all quark mass to
temperature ratios, and exhibits a first-order phase transition. Finally we
consider an isospin chemical potential and find that the spectral functions
display a resonance peak splitting, similar to the isospin meson mass splitting
observed in effective QCD models.
\end{abstract}


\maketitle
\tableofcontents

\newpage

\section{Introduction and Summary} \label{sec:introduction}

Recently in the context of gauge/gravity duality, there has been an intensive
study of the phase diagram of $\N=4$ large $N_c$ $SU(N_c)$ supersymmetric
Yang-Mills theory with added fundamental degrees of freedom, by considering the
AdS-Schwarzschild black hole background with added D7-brane probes
\cite{Babington:2003vm,Kirsch:2004km,Apreda:2005yz,Mateos:2006nu,Albash:2006ew,Hoyos:2006gb}.
There are two kinds of D7-brane probes in the black hole background: Either they
end before reaching the black hole horizon, since the $S^3$ wrapped by the
D7-brane probe shrinks to zero as in \cite{Karch:2002sh}, or they reach all the
way to the black hole horizon. The first class of embeddings is usually called
`Minkowski embeddings', while the second is referred to as `black hole
embeddings'. The parameter which parametrizes different embeddings is the
temperature normalized quark mass~$m_q/T$, which may be given in terms of the
asymptotic value~$\chi_0$ of the embedding coordinate at the AdS horizon. The
phase transition between both classes of embeddings is of first order. The
analysis of the meson spectrum shows that this phase transition corresponds to a
fundamental confinement/deconfinement transition at which the mesons melt.

Particular interest has arisen in the more involved structure of the phase
diagram when a baryon chemical potential is present \cite{Kobayashi:2006sb}. 
It was argued that for non-vanishing baryon
density, there are no embeddings of Minkowski type, and all
embeddings reach the black hole horizon. This is due to the fact that a
finite baryon density generates strings in the dual supergravity
picture which pull the brane towards the black hole. A chemical
potential for these baryons corresponds to a vev $\tilde A_0$ for the time 
component of the gauge field on the
brane. 
In the 
dual thermal $SU(N_c)$-gauge theory a baryon is composed of~$N_c$ quarks,
such that
the baryon density~$n_B$ can be directly translated into a quark density~$n_q=n_B\, N_c$.
The thermodynamic dual quantity of the quark density is the quark chemical potential~$\mu_q$. 
In the brane setup we use, the chemical potential is determined by the choice of 
quark density and by the embedding parameter~$\chi_0$.   

\begin{figure}
	\psfrag{0}{$0$}
	\psfrag{0.2}{$0.2$}
	\psfrag{0.4}{$0.4$}
	\psfrag{0.6}{$0.6$}
	\psfrag{0.8}{$0.8$}
	\psfrag{1}{$1$}
	\psfrag{muOverm}{$\mu_q/m_q$}
	\psfrag{ToverM}{$T/\bar M$}
	\psfrag{dt0}{$\tilde d = 0$}
	\psfrag{dt0002}{$\;\;\tilde d = 0.00315$}
	\psfrag{dt4}{$\tilde d = 4$}
	\psfrag{dt025}{$\tilde d = 0.25$}
        \includegraphics[width=0.9\linewidth]{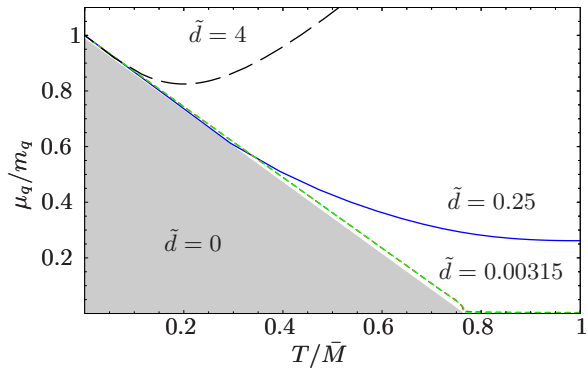}  
                \caption{
        The phase diagram for quarks: The
        quark chemical  potential~$\mu_q$ divided by the quark mass is plotted 
        versus the 
        temperature~$T$ divided by $\bar M= 2 m_q/\sqrt{\lambda}$. 
        Two different regions are displayed: The shaded region with vanishing
        baryon density and the region above the transition line with finite 
        baryon density, in which we work here. The multivalued
        region at the lower tip of the transition line is not resolved here.
        The curves are lines of equal baryon density parametrized 
        by $\tilde d=2^{5/2}n_B/(N_f\sqrt{\lambda}T^3)$.  
        The critical density~$\tilde d^*=0.00315$, at which the first order 
        phase transition between two black hole
        phases disappears, is shown as short-dashed line close to 
the transition line. It virtually coincides with the short-dashed line for~$\tilde d=0.002$.
                }
                \label{fig:phaseDiagram}
\end{figure}

Very recently, however, it was found that for a vanishing baryon number density,
there may indeed be Minkowski embeddings if a constant vev $\tilde A_0$ is
present, which does not depend on the holographic coordinate
\cite{Karch:2007br,Mateos:2007vc,Nakamura:2006xk,Nakamura:2007nx,Ghoroku:2007re}.
The phase diagram found there is sketched in figure~\ref{fig:phaseDiagram}. In
the grey shaded region, the baryon density vanishes~($n_B=0$) but temperature,
quark mass and chemical potential can be nonzero. This low temperature region
only supports Minkowski embeddings with the brane ending before reaching the
horizon. In contrast, the unshaded region supports black hole embeddings with
the branes ending on the black hole horizon. In this regime the baryon density
does not vanish~($n_B\not =0$). In this paper we exclusively explore the latter
region. At the lower tip of the line separating~$n_B=0$ from~$n_B\not=0$ in
figure~\ref{fig:phaseDiagram}, there exists also a small region of multivalued
embeddings, which are thermodynamically unstable~\cite{Mateos:2007vc}.

In the black hole phase considered here, there is a fundamental 
phase transition between different black hole 
embeddings~\cite{Kobayashi:2006sb}. 
This is a first order transition, which occurs in a region of 
the phase diagram close to the
separation line between the two regions with vanishing (gray shaded) and
non-vanishing (unshaded) baryon density. This transition disappears above a
critical value for the baryon density $n_B$ given by
\begin{equation}
        \label{eq:dcrit}
        \tilde d^*= 0.00315\,, \qquad \tilde d = 2^{5/2}n_B/(N_f\sqrt{\lambda}T^3)\,.
\end{equation}

In this paper we make use of the methods developed in the context of AdS/CFT applied
to hydrodynamics, for instance \cite{Son:2002sd,Teaney:2006nc,Kovtun:2006pf}, in
order to determine the spectral function at finite temperature and finite baryon
density. For vanishing chemical potential, a similar analysis of the spectral
functions has been performed in \cite{Myers:2007we}. It was found that the
spectrum is discrete at large quark mass, or equivalently at low
temperature.  
At low quark mass, a quasiparticle structure is seen which displays the
broadening decay width of the mesons. As the mass decreases or temperature
rises, the mesons are rendered unstable as the resonance frequencies develop
imaginary parts. Modes corresponding to such frequencies are called quasinormal.
These excitations are then dissipated in the plasma. -- Note that for
this case, there are also lattice gauge theory results \cite{Aarts:2007wj}.

In this paper we study the differences in the spectral functions with and
without chemical potential. Relating our work to the phase diagram shown
in~\cite[figure 2]{Mateos:2007vc}~(and reproduced here in
figure~\ref{fig:phaseDiagram}), we here consider the region of black hole
embeddings~(unshaded region) with nonvanishing quark density. We find that at low   
temperature to quark mass ratio, the spectrum is asymptotically
discrete and coincides with the zero-temperature supersymmetric meson
mass formula found in~\cite{Kruczenski:2003be}, which in our coordinates reads
\begin{equation}
\label{eq:massFormula}
M=\frac{L_\infty}{R^2}\,\sqrt{2(n+1)(n+2)}\,.   
\end{equation}
In \cite{Kruczenski:2003be}, $L_\infty$ denotes the asymptotic separation of the
D3- and D7-branes and $n$ counts the nodes of the embedding fluctuations. Here
we are considering s-wave modes in the Kaluza-Klein expansion of the D7 brane
probe wrapping $S^3$, so the angular quantum number~$l$ is zero. We connect the
structure of the spectra found to the phase diagram in
figure~\ref{fig:phaseDiagram}: The meson mass behavior described above occurs
close to the Minkowski region of the phase diagram, where temperature effects
are subdominant. Moreover, as a function of decreasing temperature to quark mass
ratio, the quasiparticle peaks behave differently with and without finite quark
density. As discussed in \cite{Myers:2007we}, at vanishing density~$n_B=0$ the
peak maxima move towards smaller frequencies as a function of increasing quark
mass. Here, in the case of finite quark density, we observe a similar behavior
at small quark mass. However, keeping the temperature fixed as we increase the
quark mass further, the peaks turn around at a value $m_q^{\text{turn}}$ and
move to larger and larger frequencies as the associated mesons become more
stable. Note that a turning point behavior was also observed for vanishing quark
density in the context of quasinormal modes for scalar modes of
melting mesons~\cite{Hoyos:2006gb}.

Our spectra also show that for given quark mass and temperature, lower $n$ meson
excitations can be nearly stable in the plasma, while higher $n$ excitations
remain unstable. At vanishing baryon density, the formation of
resonance peaks for higher excitations has also been observed
in~\cite{Mateos:2007yp}. We discuss the distinct behavior of
resonance peaks in section~\ref{sec:vectorResults}, including a comparison of
the observed turning points at finite baryon density with previous results.

We also calculate the quark diffusion constant $D$ and show that at finite
density, it exhibits the first-order fundamental phase transition up to the
critical density given by $\tilde d^*=0.00315$. For very large values of the
density, the diffusion constant asymptotes to $D \cdot T = 1/(2 \pi)$. 
This reflects the fact that in this case, the free quarks outnumber the quarks
bound in mesons.

As a second point we consider the case of an isospin chemical potential, on
which previous work in the holographic context has appeared in
\cite{Apreda:2005yz,Parnachev:2007bc}. In this case, two coincident D7-brane
probes are considered. In particular we extend the results of our previous paper
\cite{Erdmenger:2007ap}, in which we calculated the retarded Green function and
diffusion coefficient at finite $SU(2)$ isospin chemical potential for the flat
embedding $m_q=0$. In this previous work we also restricted to the case of
constant vev for the non-Abelian gauge field $A^3_0$, where $3$ is the flavor
and $0$ the Lorentz index. This means that we chose $A^3_0$ to be independent of
the $AdS$ radial direction. In this case we found a non-analytic frequency
dependence of the Green functions and the diffusion coefficient. Here we extend
this work to the case of non-vanishing quark mass, leading to
non-trivial D7 embeddings, and to the case of radially varying gauge field
component $A^3_0$. We find that spectral functions quantitatively deviate from
the baryonic background case. Additionally, a splitting of quasi-particle
resonances is observed, which depends on the magnitude of the chemical
potential.

This paper is organized as follows. In the following section~\ref{sec:setup}, we
introduce the gravity background, field and brane configuration, used for the
subsequent calculations. We also sketch the method to obtain retarded real-time
correlators of thermal field theories from supergravity calculations. In
section~\ref{sec:baryonMu} we discuss the spectral functions and diffusion
behavior of fundamental matter at finite baryon density. For matter with isospin
chemical potential, the same analysis is carried out in
section~\ref{sec:isospinMu}. The results are briefly summarized in
section~\ref{sec:conclusion}.

\section{Holographic setup and thermodynamics}
\label{sec:setup}

\subsection{Background and brane configuration}
\label{sec:backAndBranes}

We consider asymptotically $AdS_5\times S^5$ space-time which arises 
as the near horizon \marginpar{test}
limit of a stack of $N_c$ coincident D3-branes. More precisely, our background
is an $AdS$ black hole, which is the geometry dual to a field theory at finite
temperature (see e.g.~\cite{Policastro:2002se}).
We make use of the coordinates of \cite{Kobayashi:2006sb} to write this
background in Minkowski signature as
\begin{equation}
\begin{split}
\label{eq:AdSBHmetric}
\dd s^2 =\; &\frac{1}{2} \left(\frac{\varrho}{R}\right)^2
\left( -\frac{f^2}{\tilde f}\dd t^2 + \tilde{f} \dd \bm{x}^2 \right)\\
 & + \left(\frac{R}{\varrho}\right)^2\left( \dd\varrho^2 +\varrho^2 \dd\Omega_5^2
  \right)  ,
\end{split}
\end{equation}
with the metric $\mathrm d \Omega_5^2$ of the unit $5$-sphere, where
\marginpar{test}
\begin{equation}
\begin{split}
\label{eq:metricDefinitions}
f(\varrho)=1-\frac{\varrho_H^4}{\varrho^{4}},\qquad\tilde f(\varrho)=1+\frac{\varrho_H^4}{\varrho}^{4},\\
R^4=4\pi g_s N_c  {\alpha'}^2,\qquad \varrho_H = T \pi R^2.
\end{split}
\end{equation}
Here $R$ is the $AdS$ radius, $g_s$ is the string coupling constant, $T$ the
temperature, $N_c$ the number of colors. In the following some equations may be
written more conviniently in terms of the dimensionless radial coordinate
$\rho=\varrho/\varrho_H$, which covers a range from $\rho=1$ at the event
horizon to $\rho \to \infty$, representing the boundary of $AdS$ space.

Into this ten-dimensional space-time we embed $N_f$ coinciding D7-branes,
hosting flavor gauge fields $A_\mu$. The embedding we choose lets the D7-branes
extend in all directions of $AdS$ space and, in the limit $\rho \rightarrow
\infty$, wraps an $S^3$ on the $S^5$. It is convenient to write the D7-brane
action in coordinates where
\begin{equation}
 \dd \varrho^2 +\varrho^2 \dd \Omega_5^2 = \dd \varrho^2 + \varrho^2 ( \dd \theta^2 
 + \cos^2 \theta \dd \phi^2 + \sin^2 \theta \dd \Omega_3^2) ,
\end{equation}
with $0\leq\theta<\pi/2$. From the viewpoint of ten dimensional Cartesian $AdS_5
\times S^5$, $\theta$ is the angle between the subspace spanned by
the 4,5,6,7-directions, into which the D7-branes extend perpendicular to the
D3-branes, and the subspace spanned by the 8,9-directions, which are transverse
to all branes.

Due to 
the symmetries of this background, the embeddings depend only on the
radial coordinate $\rho$. Defining $\chi \equiv \cos \theta$, the embeddings of
the D7-branes are parametrized by the functions~$\chi(\rho)$. They describe
the location of the D7-branes in $8,9$-directions. Due to our choice of the
gauge field fluctuations in the next subsection, the remaining three-sphere
in this metric will not play a prominent role.

The metric induced on the D7-brane probe is then given by
\begin{equation}
\label{eq:inducedMetric} 
\begin{split}
\dd s^2 =\; & \hphantom{\:\,+}\frac{1}{2} \left(\frac{\varrho}{R}\right)^2 \left(-\frac{f^2}{\tilde f}\,\dd t^2+
  \tilde f\, \dd \bm{x}^2 \right) \\
  & +  
  {\left(\frac{R}{\varrho}\right)^2  \frac{1-\chi^2+\varrho^2{\chi'}^2}{1-\chi^2}\, \dd \varrho^2} 
  \\ & +R^2(1-\chi^2) \dd \Omega_3^2.  
\end{split}
\end{equation}
Here and in what follows we use a prime to denote a derivative with respect to
$\varrho$ (resp.\ to $\rho$ in dimensionless equations). The symbol $\sqrt{-g}$
denotes the square root of the determinant of the induced metric on the
D7-brane, which is given by
\begin{equation}
        \sqrt{-g} = \varrho^3 \frac{f \tilde f}{4}\, (1-\chi^2) \sqrt{1-\chi^2 + \varrho^2 {\chi'}^2}.
\end{equation}

The table below gives an overview of the indices we use to refer to
certain directions and subspaces.

\begin{center}
       \includegraphics[height=5.5\baselineskip]{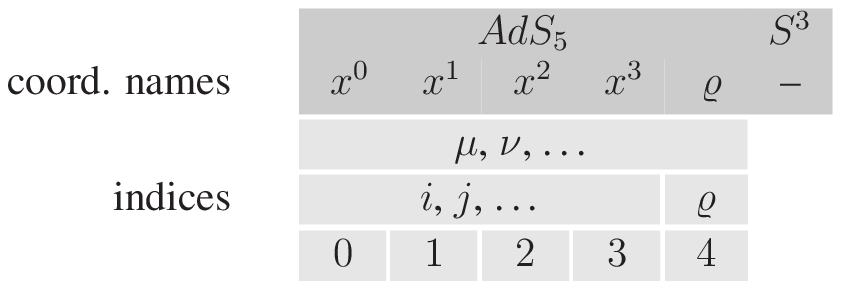}
\end{center}

The background geometry described so far is dual to thermal $\mathcal N = 4$
supersymmetric $SU(N_c)$ Yang-Mills theory with $N_f$ additional $\mathcal N =
2$ hypermultiplets. These hypermultiplets arise from the lowest excitations of
the strings stretching between the D7-branes and the background-generating
D3-branes. The particles represented by the fundamental fields of the $\mathcal
N = 2$ hypermultiplets model the quarks in our system. Their mass $m_q$ is given
by the asymptotic value of the separation of the D3- and D7-branes. In the
coordinates used here we write \cite{Myers:2007we}
\begin{equation}
\label{eq:quarkMass}
\frac{2m_q}{\sqrt{\lambda} T} = \frac{\bar M}{T} = \lim_{\rho \to \infty} \rho\, \chi(\rho) = m,
\end{equation}
where we introduced the dimensionless scaled quark mass $m$.

In addition to the parameters incorporated so far, we aim for a description of
the system at finite chemical potential $\mu$ and baryon density $n_B$. In field
theory, a chemical potential is given by a nondynamical time component of the
gauge field. In the gravity dual, this is obtained by introducing a
$\rho$-dependent gauge field component $\bar A_0(\rho)$ on the D7 brane probe.
For now we consider a baryon chemical potential which is obtained from the
$U(1)$ subgroup of the flavor symmetry group. The sum over flavors then yields a
factor of $N_f$ in front of the DBI~action written down below.

The value of the chemical potential $\mu$ in the dual field theory is then
given by
\begin{equation}
\label{eq:chemPotLimit}   
\mu = \lim_{\rho\to\infty} \bar A_0(\rho) = \frac{\varrho_H}{2\pi\alpha'} \tilde\mu,
\end{equation}
where we introduced the dimensionless quantity $\tilde\mu$ for convenience. We
apply the same normalization to the gauge field and distinguish the
dimensionful quantity $\bar A$ from the dimensionless $\tilde A_0=\bar
A_0\,(2\pi\alpha')/\varrho_H$.

The action for the probe branes' embedding function
and gauge fields on the branes is
\begin{equation}
        \label{eq:dbiAction}
        S_{\text{DBI}} = -N_f\, T_{\text{D7}}\int\!\! \dd^8 \xi\; \sqrt{| \det ( g + \tilde F )|}.
\end{equation}
Here $g$ is the induced metric \eqref{eq:inducedMetric} on the brane,
$\tilde F$ is the field strength tensor of the gauge fields on the brane and $\xi$ are
the branes' worldvolume coordinates. $T_{\text{D7}}$ is the brane tension and
the factor $N_f$ arises from the trace over the generators of the
symmetry group under consideration. For finite baryon density, this factor will be different 
from that at finite isospin density.

In~\cite{Kobayashi:2006sb}, the dynamics of this system of branes and gauge
fields was analyzed in view of describing phase transitions at finite baryon
density. Here we use these results as a starting point which gives the
background configuration of the brane embedding and the gauge field values at
finite baryon density. To examine vector meson spectra, we will then investigate
the dynamics of fluctuations in this gauge field background.

In the coordinates introduced above, the action $S_{\text{DBI}}$ for the
embedding $\chi(\rho)$ and the gauge fields' field strength $F$ is obtained by
inserting the induced metric and the field strength tensor into
\eqref{eq:dbiAction}. As in \cite{Kobayashi:2006sb}, we get
\begin{multline}
\label{eq:actionEmbeddingsAt} 
S_{\text{DBI}}=-N_f T_\mathrm{D7} \varrho_H^3 \int\!\! \dd^8 \xi\; \frac{\rho^3}{4} f
\tilde{f} (1-\chi^2)\\ 
\times\sqrt{1-\chi^2+\rho^2 
{\chi'}^2-2 \frac{\tilde{f}}{f^2}(1-\chi^2) \tilde F_{\rho 0}^2} \; , 
\end{multline}
where $\tilde F_{\rho 0}=\partial_\rho \tilde A_0$ is the field strength on the
brane. $\tilde A_0$ depends solely on $\rho$.

\begin{figure}
	\psfrag{m}{$m$}
	\psfrag{chi0}{$\chi_0$}
	\psfrag{dt0}{$\tilde d = 0$}
	\psfrag{dt0002}{$\tilde d = 0.002$}
	\psfrag{dt01}{$\tilde d = 0.1$}
	\psfrag{dt025}{$\tilde d = 0.25$}
	\includegraphics[width=0.9\linewidth]{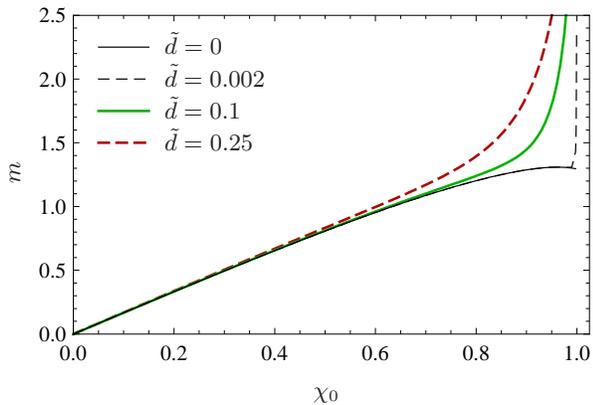}   
	\caption{The dependence of the scaled quark mass~$m=2 m_q/\sqrt{\lambda} T$
		on the horizon value~$\chi_0=\lim_{\rho\to1}\chi$ of the embedding.}
	\label{fig:mOfChi0}   
\end{figure}

According to \cite{Kobayashi:2006sb}, the equations of motion for the background
fields are obtained after Legendre transforming the
action~\eqref{eq:actionEmbeddingsAt}. Varying this Legendre transformed action
with respect to the field~$\chi$ gives the equation of motion for the
embeddings~$\chi(\rho)$,
\begin{equation}
\begin{aligned}
\label{eq:eomChi}
 & \partial_\rho\left[\frac{\rho^5 f \tilde{f} (1-\chi^2)
{\chi'}}{\sqrt{1-\chi^2+\rho^2{\chi'}^2}} \sqrt{1 +
\frac{8 \tilde{d}^2}{\rho^6 \tilde{f}^3 (1-\chi^2)^3}}\right] \\
= & - \frac{\rho^3 f \tilde{f} \chi }{\sqrt{1-\chi^2+\rho^2{\chi'}^2}}
\sqrt{1 +\frac{8 \tilde{d}^2}{\rho^6 \tilde{f}^3 (1-\chi^2)^3}}\\
 & \times\left[3 (1-\chi^2) +2 \rho^2 {\chi'}^2 -24 \tilde{d}^2
\frac{1-\chi^2+\rho^2{\chi'}^2}{\rho^6 \tilde{f}^3 (1-\chi^2)^3+8 \tilde{d}^2}
\right] .
\end{aligned}
\end{equation}
The dimensionless quantity $\tilde d$ is a constant of motion. It is related to
the baryon number density $n_B$ by \cite{Kobayashi:2006sb}
\begin{equation}
        \label{eq:nbfromdtilde}
        n_B = \frac{1}{2^{5/2}} N_f \sqrt{\lambda} T^3 \tilde d.
\end{equation}
Below,  equation \eqref{eq:eomChi} will be solved numerically for different
initial values~$\chi_0$ and~$\tilde d$. The boundary conditions used are
\begin{equation}
\chi(\rho=1)=\chi_0,\qquad  \partial_\rho \chi(\rho) \Big|_{\rho=1}=0 .
\end{equation}
The quark mass $m$ is determined by $\chi_0$. It is zero for $\chi_0=0$ and
tends to infinity for $\chi_0 \to 1$. Figure~\ref{fig:mOfChi0} shows the
dependence of the scaled quark mass~$m=2 m_q/\sqrt{\lambda}T$ on the starting value~$\chi_0$
for different values of the baryon density parametrized by~$\tilde d\propto n_B$.
In general, a small (large)~$\chi_0$ is equivalent to a small (large) quark mass.
For~$\chi_0<0.5$, $\chi_0$ can be viewed as being proportional
 to the large quark masses. At
larger~$\chi_0$ for vanishing~$\tilde d=0$, the quark mass reaches a finite value.
In contrast, at finite baryon density, if~$\chi_0$ is close to
$1$, the mass rapidly increases when increasing~$\chi_0$ further. In
embeddings with a phase transition, there exist more than one embedding for one
specific mass value. In a small regime close to~$\chi_0=1$, there are more than
one possible value of~$\chi_0$ for a given~$m$. So in this small region,
$\chi_0$ is not proportional to~$m_q$.   

\begin{figure}
	\psfrag{chi}[lc]{\small$\chi$}
	\psfrag{rho}[lc]{\small$\rho$}
	\psfrag{r}[lc]{\small$r$}
	\psfrag{L}[cB]{\small$L$}
	\psfrag{At}[cB]{\small$\tilde A_0$}
	\psfrag{Atin10ToMinus4}[cB]{\small$\tilde A_0/10^{-4}$}
	\psfrag{L}[cB]{\small$L$}
	\psfrag{0}[lc]{\tiny $0$}
	\psfrag{0.2}[lc]{\tiny $0.2$}
	\psfrag{0.4}[lc]{\tiny $0.4$}
	\psfrag{0.5}[lc]{\tiny $0.5$}
	\psfrag{0.6}[lc]{\tiny $0.6$}
	\psfrag{0.8}[lc]{\tiny $0.8$}
	\psfrag{1}[lc]{\tiny $1$}
	\psfrag{1.5}[lc]{\tiny $1.5$}
	\psfrag{2.5}[lc]{\tiny $2.5$}
	\psfrag{2}[lc]{\tiny $2$}
	\psfrag{3}[lc]{\tiny $3$}
	\psfrag{4}[lc]{\tiny $4$}
	\psfrag{5}[lc]{\tiny $5$}
\centering
\includegraphics[width=0.49\linewidth]{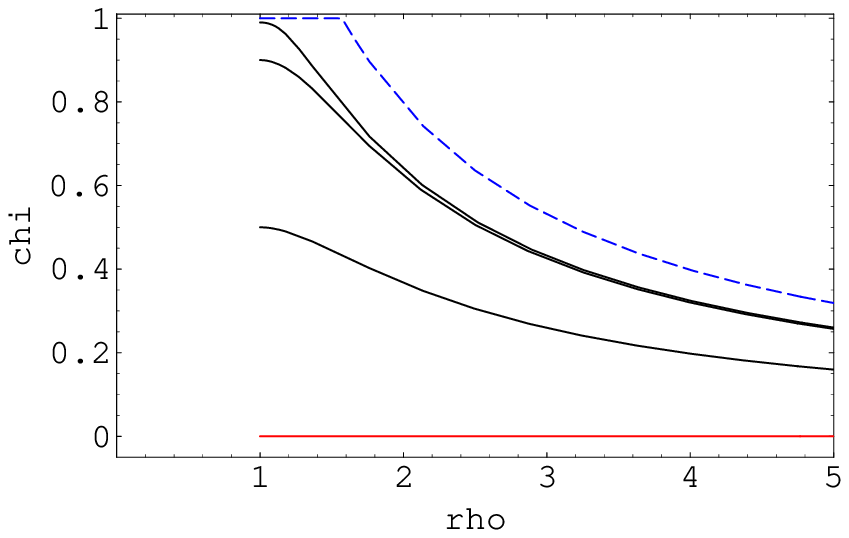}
\hfill
\includegraphics[width=0.49\linewidth]{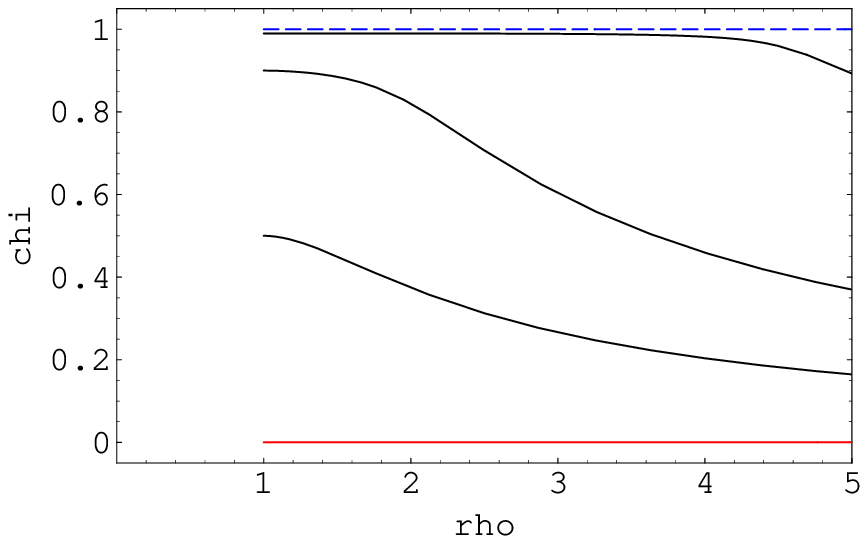}
\vfill\vspace{5pt}  
\includegraphics[width=0.49\linewidth]{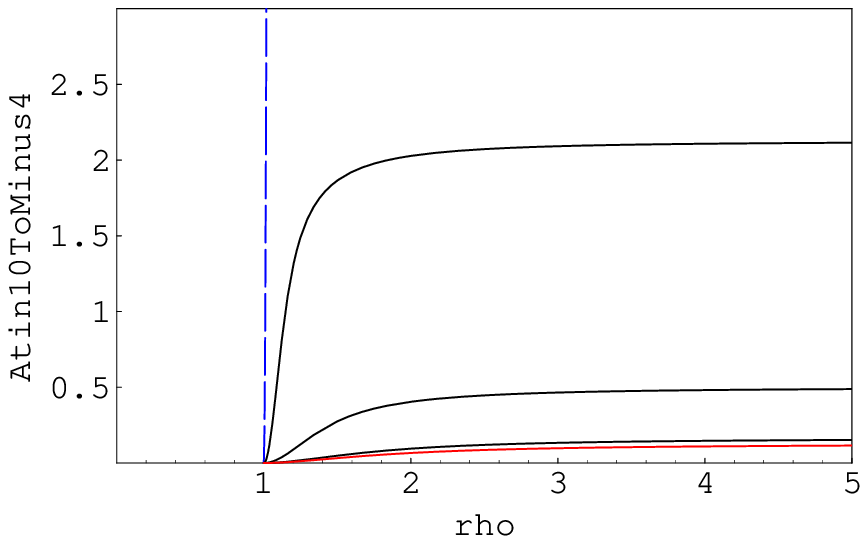}
\hfill
\includegraphics[width=0.49\linewidth]{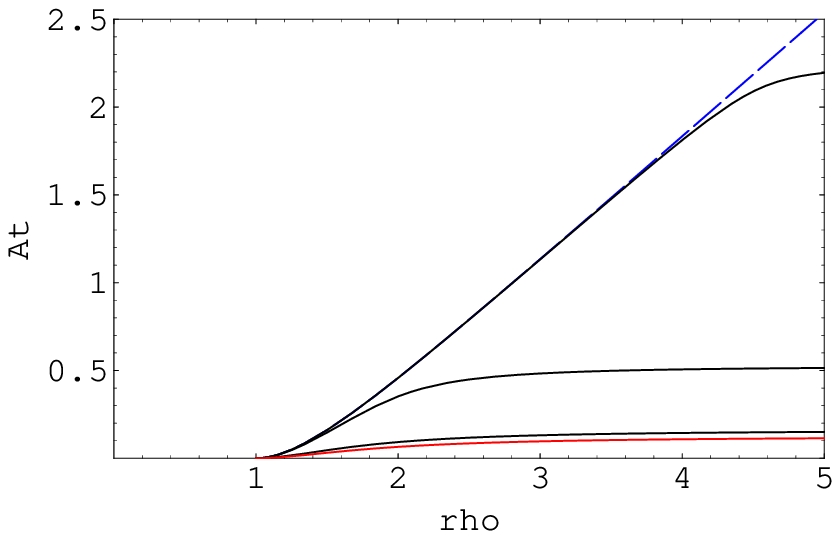}
\vfill\vspace{5pt}  
\includegraphics[width=0.49\linewidth]{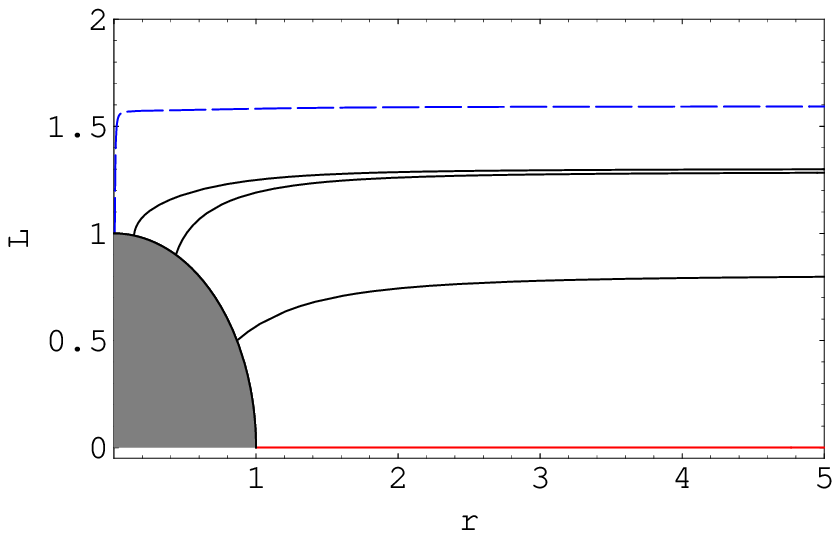}
\hfill
\includegraphics[width=0.49\linewidth]{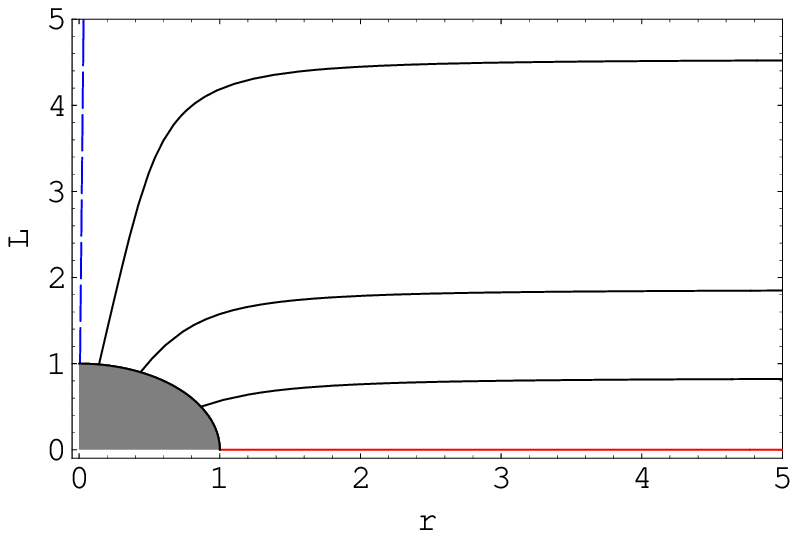}
\caption{\label{fig:backgAt}
        The three figures of the left column show the embedding function~$\chi$
        versus the radial coordinate~$\rho$, the corresponding background gauge 
        fields~$\tilde A_0$ and the distance~$L=\rho\,\chi$ between the D3 and
		the D7-branes at $\tilde d=10^{-4}/4$.
		$L$ is plotted versus $r$, given by $\rho^2 = r^2 + L^2$.
		In the right column, the same three quantities are
        depicted for~$\tilde d=0.25$. 
        The five curves in each plot correspond to parametrizations of the 
        quark mass to temperature ratio with~$\chi_0=\chi(1)=0,\,0.5,\,0.9,\,0.99$~(all solid)  
	and $0.99998$~(dashed) from bottom up.
        These correspond to scaled quark 
        masses~$m=2 m_q/T\sqrt{\lambda}=0,\,0.8089,\,1.2886,\,1.3030,\,1.5943$ in
        the left plot and to~$m=0,\,0.8342,\,1.8614,\,4.5365,\,36.4028$ on the right.  
	The curves on the left
        exhibit~$\mu\approx 10^{-4}$. Only the upper most curve    
        on the left at~$\chi_0=0.99998$ develops 
        a large chemical potential of~$\mu=0.107049$. In the right column 
	curves correspond to chemical potential values~$\mu=0.1241,\,0.1606,\,
	0.5261,\,2.2473,\,25.3810$~from bottom up.  
        }
\end{figure}

The equation of motion for the background gauge field $\tilde A$ is 
\begin{equation}
\label{eq:eomD}
 \partial_\rho\tilde{A}_0 = 2 \tilde{d}
\frac{f \sqrt{1-\chi^2+\rho^2 {\chi'}^2}}
{\sqrt{\tilde{f}(1-\chi^2) [\rho^6 \tilde{f}^3 (1-\chi^2)^3+8 \tilde{d}^2]}}. 
\end{equation}
Integrating both sides of the equation of
motion from
$\rho_{\text{H}}$ to some $\rho$, and respecting the boundary condition $\tilde
A_0(\rho=1)=0$ \cite{Kobayashi:2006sb}, we obtain the full
background gauge field
\begin{equation}
\label{eq:backgroundAt}
 \tilde{A}_0(\rho) = 2\tilde{d}
\int\limits_{\rho_H}^\rho\!\! \dd\rho\;
\frac{f\sqrt{1-\chi^2+\rho^2{\chi'}^2}}{\sqrt{\tilde{f}
(1-\chi^2)[\rho^6 \tilde{f}^3 (1-\chi^2)^3 +8 \tilde{d}^2]}}\,.
\end{equation}
Recall that the chemical potential of the field theory is given by
$\lim_{\rho\to\infty}\tilde A_0(\rho)$ and thus can be obtained from the formula
above.  Examples for the functional behavior of $A_0(\rho)$ are shown in
figure~\ref{fig:backgAt}. Note that at a given baryon density $n_B\neq 0$ there
exists a minimal chemical potential which is reached in the limit of massles
quarks.

The asymptotic form of the fields $\chi(\rho)$ and $A_0(\rho)$ can be found from
the equations of motion in the boundary limit~$\rho\to\infty$,
\begin{align}
\label{eq:boundaryAt}
\bar A_0 & =\mu - \frac{\tilde d}{\rho^2} \frac{\varrho_H}{2\pi\alpha'} + \cdots, \\
\label{eq:boundaryChi}
\chi & =\frac{m}{\rho}+\frac{c}{\rho^3}+ \cdots .
\end{align}
Here $\mu$ is the chemical potential, $m$ is the dimensionless quark mass
parameter given in \eqref{eq:quarkMass}, $c$ is related to the quark condensate
(but irrelevant in this work) and $\tilde d$ is related to the baryon number
density as stated in \eqref{eq:nbfromdtilde}. See also figure~\ref{fig:backgAt}
for this asymptotic behavior.
The $\rho$-coordinate runs from the horizon value~$\rho=1$ to the boundary
at~$\rho=\infty$. In most of this range, the gauge field is almost constant
and reaches its asymptotic value, the chemical potential~$\mu$, at
$\rho\to\infty$. Only near the horizon the field drops rapidly to zero. For
small~$\chi_0\to 0$, the curves asymptote to the lowest~(red) curve. So
there is a minimal chemical potential for fixed baryon density in this
setup. At small baryon density~($\tilde d \ll 0.00315$) the embeddings
resemble the Minkowski and black hole embeddings known from the case 
without a chemical potential. Only a thin spike always reaches down to 
the horizon. 
 
In the setup described in this section we restrict ourselves to the regime of so
called `black hole embeddings' which are those embeddings ending on the horizon
of the black hole, opposed to `Minkowski embeddings' , which would reach
$\rho=0$ without touching the horizon. The black hole embeddings we use for
this work~(see figure~\ref{fig:backgAt}) are not capable of describing matter in all possible phases. In fact
we are able to cover the regime of fixed $n_B > 0$ and thus examine thermal
systems in the canonical ensemble at finite baryon density. For a detailed
discussion of this aspect see~\cite{Karch:2007br,Mateos:2007vc}.

\subsection{Holographic spectral functions}
\label{sec:specFuncs}

Spectral functions contain information about the quasiparticle spectrum of a
given theory. Recently, methods were developed to compute spectral functions
from the holographic duals of strongly coupled finite temperature gauge
theories. In this work we extend these results to investigate the
quasiparticle spectrum corresponding to vector mesons in the limit of vanishing
spatial momentum. Therefore, we analyze the holographic dual to spectral
functions for thermal $\mathcal N = 4$ supersymmetric $SU(N_c)$
Yang-Mills theory with $ N_f$ fundamental degrees of freedom (quarks) at
finite baryon density and finite chemical potential. We compute   
the spectral densities for the flavor current $J$, which is dual to
the fluctuations $A$ of the flavor gauge field on the supergravity side.  

Within field theory, the spectral function $\R(\omega, \bm q)$ of some operator
$J(x)$ is defined via the imaginary part of the retarded Green
function $G^R$ as follows
\begin{equation}
        \label{eq:specDensity}
        \R(\omega, \bm q) = -2 \:\mathrm{Im} G^R(\omega, \bm q) \, ,
\end{equation}
where Energy $\omega$ and spatial momentum $\bm q$ may be written in a four
vector $\vec k=(\omega,\bm q)$ and the Green function $G^R$ may be written as
\begin{equation}
        G^R(\omega, \bm q) = -i \int\!\! \dd^4x\:  e^{i\:\vec{k}\vec{x}}\, \theta(x^0) \left< \left[ J(\vec{x}), J(0) \right] \right>
\end{equation}
One may find singularities of $G^R(\omega, \bm q)$ in the lower half of the complex
$\omega$-plane, including hydrodynamic poles of the retarded real-time Green function. Consider for example
\begin{equation}
        \label{eq:quasiNormalModes}
        G^R = \frac{1}{\omega-\omega_0 + i \Gamma}.
\end{equation}
These poles emerge as peaks in the spectral densities,
\begin{equation}
        \R = \frac{2\, \Gamma}{(\omega-\omega_0)^2+\Gamma^2}\,,
\end{equation}
located at $\omega_0$ with a width given by $\Gamma$. These peaks are
interpreted as quasi-particles if their lifetime $1/\Gamma$ is considerably
long, i.e.\ if $\Gamma \ll \omega_0$.

In this paper we use the gauge/gravity duality prescription of \cite{Son:2002sd}
for calculating Green functions in Minkowski space-time. For further reference,
we outline this prescription briefly in the subsequent. Starting out from a
classical supergravity action~$S_{\rm cl}$ for the gauge field $A$, according to
\cite{Son:2002sd} we extract the function~$B(\rho)$~(containing metric factors
and the metric determinant) in front of the kinetic term~$(\partial_\rho A)^2$,
\begin{equation}
\label{eq:classicalAction}
S_{\mathrm{cl}}= \,\int\!\! \mathrm d \rho \, \mathrm d^4x\, \, 
  B(\rho)\,(\partial_\rho A)^2\, +\, \dots\,.
\end{equation}
Then we perform a Fourier transformation and solve the linearized equations of
motion for the fields $A$ in momentum space. The solutions in general are
functions of all five coordinates in Anti-de~Sitter space.  Near the boundary
we may separate the radial behavior from the boundary dynamics by writing
\begin{equation}
\label{eq:rhoIs0Boundary}
A(\rho,\vec k) = f(\rho,\vec k)\, A^{\text{bdy}}(\vec k) \, ,
\end{equation}
where~$A^{\text{bdy}}(\vec k)$ is the value of the supergravity field at the
boundary of $AdS$ depending only on the four flat boundary coordinates. Thus by
definition we have~$f(\rho,\vec k)|_{\rho\to\infty}=1$. Then the retarded
thermal Green function is given by
\begin{equation}
\label{eq:retardedThermalGreen}
G^R(\omega, {\bm q})\,=\, \left. 2\, B(\rho)\, f(\rho,-\vec k)\,
  \partial_\rho\,f(\rho,\vec k)\right|_{\rho\to\infty} \, .
\end{equation}

The thermal correlators obtained in this way display hydrodynamic properties,
such as poles located at complex frequencies. They are used to compute the
spectral densities \eqref{eq:specDensity}. We are going to compute the functions
$A(\rho,k)$ numerically in the limit of vanishing spatial momentum $\bm q\to 0$.
The functions $f(\rho,\vec k)$ are then obtained by dividing out the boundary
value $A^{\text{bdy}}(\vec k) = \lim_{\rho\to\infty} A(\rho,\vec k)$.
Numerically we obtain the boundary value by computing the solution at a fixed
large $\rho$.

\section[Spectral functions at finite baryon density]{
Spectral functions at finite baryon density}
\label{sec:baryonMu}

\subsection{Baryon diffusion}
\label{sec:baryonDiff}

In this section we calculate the baryon diffusion coefficient and its dependence
on the baryon density. As discussed in \cite{Mateos:2007vc}, the baryon density
affects the location and the presence of the fundamental phase transition
between two black hole embeddings observed in~\cite{Kobayashi:2006sb}. This
first order transition is present only very close to the separation line between
the regions of zero and non-zero baryon density shown in
figure~\ref{fig:phaseDiagram}.

We show that this fundamental phase transition may also be seen in the diffusion
coefficient for quark diffusion. In order to compute the diffusion using
holography, we use the membrane paradigm approach developed
in~\cite{Kovtun:2003wp} and extended in~\cite{Myers:2007we}. This method allows
to compute various transport coefficients in Dp/Dq-brane setups from the metric
coefficients. The resulting formula for our background is the same as
in~\cite{Myers:2007we},
\begin{equation}
\label{eq:diffusionConstant} 
D = \left .\frac{
        \sqrt{-g}}{
        g_{11}\sqrt{-g_{00} g_{44}}}\right |_{\rho=1}
\int\!\! \mathrm{d}\rho\;\frac{-g_{00} g_{44}}{\sqrt{-g}} .   
\end{equation}

\begin{figure}
        \includegraphics[width=.9\linewidth]{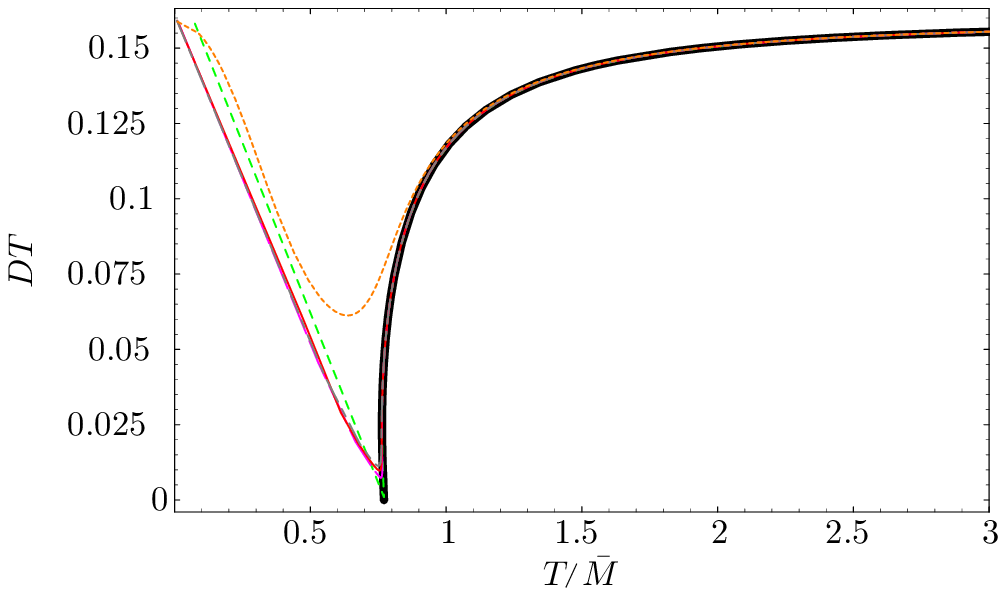}
\vfill
        \includegraphics[width=.9\linewidth]{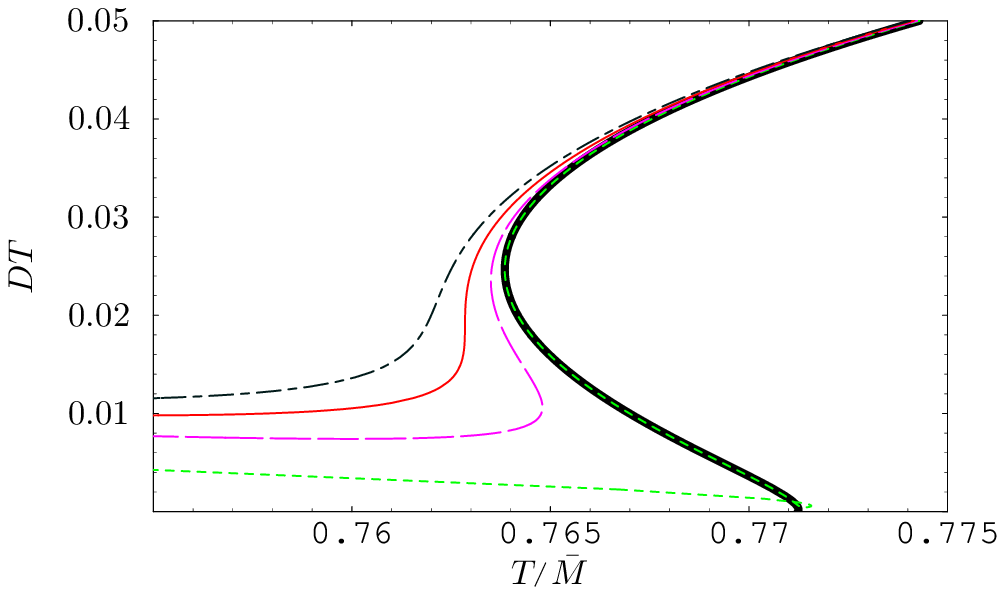}
        \caption{
                The diffusion coefficient times temperature is plotted against
                the mass-scaled temperature for diverse baryon densities
                parametrized by~$\tilde d=0.1$~(uppermost line in upper plot, not
		visible in lower plot),
                $0.004,$~(long-dashed), $0.00315$~(thin solid),
                $0.002$~(long-short-dashed), $0.000025$~(short-dashed) and
                $0$~(thick solid). The finite baryon density lifts the curves at
                small temperatures. Therefore the diffusion constant never
                vanishes but is only minimized near the phase transition. The
                lower plot zooms into the region of the transition. 
                The phase transition vanishes above a critical
                value~$\tilde d^*=0.00315$. The position of the
                transition shifts to smaller~$T/\bar M$, as~$\tilde d$ is
                increased towards its critical value.
		}
        \label{fig:diffusionConstants}
\end{figure}

The dependence of $D$ on the baryon density and on the quark mass originates
from the dependence of the embedding $\chi$ on these variables. The results
for~$D$ are shown in figure~\ref{fig:diffusionConstants}. The thick solid line
shows the diffusion constant at vanishing baryon density found in
in~\cite{Myers:2007we}, which reaches $D=0$ at the fundamental phase transition.
Increasing the baryon density, the diffusion coefficient curve is lifted up for
small temperatures, still showing a phase transition up to the critical
density $\tilde d^*=0.00315$. This is the same value as found
in~\cite{Kobayashi:2006sb} in the context of the phase transition of the quark
condensate.

The diffusion coefficient never vanishes for finite density. Both in the limit
of $T/\bar M\to 0$ and $T/\bar M\to \infty$, $D\cdot T$ converges to $1/2\pi$
for all densities, i.e.\ to the same value as for vanishing baryon density, as
given for instance in \cite{Kovtun:2003wp} for R-charge diffusion. At the phase
transition, the diffusion constant develops a nonzero minimum at finite baryon
density. Furthermore, the location of the first order phase transition moves to
lower values of $T/\bar M$ while we increase~$\tilde d$ towards its critical
value.

In order to give a physical explanation for this behavior, we focus on the case
without baryon density first. We see that the diffusion coefficient vanishes at
the temperature of the fundamental deconfinement transition. This is simply due
to the fact that at and below this temperature, all charge carriers are bound
into mesons not carrying any baryon number.

For non-zero baryon density however, there is a fixed number of charge carriers
(free quarks) present at any finite temperature. This implies that the diffusion
coefficient never vanishes. Switching on a very small baryon density, even below
the phase transition, where most of the quarks are bound into mesons, by
definition there will still be a finite amount of free quarks. By increasing the
baryon density, we increase the amount free quarks, which at some point
outnumber the quarks bound in mesons. Therefore in the large density limit the
diffusion coefficient approaches $D^0 = 1/(2\pi T)$ for all values of $T/\bar
M$, because only a negligible fraction of the quarks is still bound in this
limit.

Note that as discussed in  \cite{Karch:2007br,Kobayashi:2006sb,Mateos:2007vc}
there exists a region in the $(n_B,T)$ phase diagram at small $n_B$ and $T$
where the embeddings are unstable. In figure~\ref{fig:diffusionConstants}, this
corresponds to the region just below the phase transition at small baryon
density. This instability disappears for large $n_B$.

\subsection{Vector mesons in the black hole phase}
\label{sec:vectorMesonsBH}

\subsubsection{Application of calculation method}
\label{sec:vectorCalc}

We now compute the spectral functions of flavor currents at finite baryon
density~$n_B$, chemical potential~$\mu$ and temperature in the `black hole
phase'. As black hole phase the authors of~\cite{Karch:2007br} denote the phase
of matter which has nonzero baryon density. Compared to the limit of vanishing
chemical potential treated in~\cite{Myers:2007we}, we discover a qualitatively
different behavior of the finite temperature oscillations corresponding to
vector meson resonances.

To obtain the spectral functions, we compute the correlations of flavor gauge
field fluctuations $A_\mu$ about the background given
by~\eqref{eq:actionEmbeddingsAt}, denoting the full gauge field by
\begin{equation}
\label{eq:hatA}  
        \hat A_\mu(\rho, \vec{x}) = \delta^0_\mu \tilde A_0(\rho) + A_\mu(\vec{x},\rho)\,.
\end{equation}
According to section~\ref{sec:backAndBranes}, the background field has a
non-vanishing time component, which depends solely on $\rho$. The fluctuations
in turn are gauged to have non-vanishing components along the Minkowski
coordinates $\vec x$ only and only depend on these coordinates and on $\rho$.
Additionally they are assumed to be small, so that it suffices to consider their
linearized equations of motion. At this point we neglect the
fluctuation of the scalar and pseudoscalar modes and their coupling to the
vector fluctuations. In fact there is no such coupling in the limit of
vanishing spatial momentum, which we restrict to below.

The resulting equations of motion are obtained from the
action \eqref{eq:dbiAction}, where we introduce small fluctuations $A$ by
setting $\tilde F_{\mu\nu} \to \hat F_{\mu\nu} = 2\, \partial_{[\mu}\hat
A_{\nu]}$ with $\hat A = \tilde A + A$. The background gauge field~$\tilde A$ is
given by~\eqref{eq:eomD}. The fluctuations now propagate on a background $G$
given by
\begin{equation}
	G {\;\equiv\;} g + \tilde F,
\end{equation}
and their dynamics is determined by the Lagrangian
\begin{equation}
	\mathcal{L} = \sqrt{\left| \det ( G + F )\right|},
\end{equation}
with the fluctuation field strength $F_{\mu\nu} = 2\partial_{[\mu}A_{\nu]}$.
Since the fluctuations and their derivatives are chosen to be small, we consider
their equations of motion only up to linear order, as derived from the part of
the Lagrangian $\mathcal{L}$ which is quadratic in the fields and their
derivatives. Denoting this part by $\mathcal{L}_2$, we get
\begin{equation}
\begin{split}
	\mathcal{L}_2 = & {-\frac{1}{4}} \sqrt{\left|\det G\right|}\;\\
					& \times \left( G^{\mu\alpha}G^{\beta\gamma} F_{\alpha\beta}F_{\gamma\mu} {-\frac{1}{2}\, G^{\mu\alpha}G^{\beta\gamma}F_{\mu\alpha}F_{\beta\gamma}}\right).
\end{split}
\end{equation}
Here and below we use upper indices on $G$ to denote elements of $G^{-1}$. The
equations of motion for the components of $A$ are
\begin{equation}
\label{eq:eomFluct}
\begin{split}
	0= & \partial_\nu\Bigg[  \sqrt{\left|\det G\right|} \\
                        & \times \left( G^{\mu\nu}G^{\sigma\gamma}-G^{\mu\sigma}G^{\nu\gamma} {- G^{[\nu \sigma]} G^{\gamma\mu}} \right) \partial_{[\gamma}A_{\mu ]} \Bigg].
\end{split}
\end{equation}
The terms of the corresponding on-shell action at the $\rho$-boundaries are
(with $\rho$ as an index for the coordinate $\rho$, not summed)
\begin{multline}
	S^{\text{on-shell}}_{\text{D7}} =  \varrho_H \pi^2 R^3 N_f T_{\text{D7}} \int\!\! \dd^4 x \sqrt{\left|\det G \right|}\\
 \times \Big( \left( G^{04}\right)^2 A_0 \partial_\rho A_0 - G^{44} G^{ik} A_i \partial_\rho A_k \\ { - A_0 G^{4 0} \mathrm{tr}(G^{-1}F)}\Big)\Bigg|^{\rho_B}_{\varrho_H}.
\end{multline}
Note that on the boundary $\rho_B$ at $\rho\to\infty$, the background matrix $G$
reduces to the induced D7-brane metric $g$. Therefore, the analytic expression
for boundary contributions to the on-shell action is identical
to the one found in \cite{Myers:2007we}.
There, the coordinates in Minkowski directions were chosen
such that the fluctuation four vector $\vec k$ exhibits only one non vanishing
spatial component, e.\,g.~in $x$-direction as $\vec{k}=(\omega,q,0,0)$. Then
the action was expressed in terms of the gauge invariant field component
combinations
\begin{equation}
E_x=\omega A_x+ q A_0,\qquad E_{y,z}=\omega A_{y,z}\, .
\end{equation}
In the case of vanishing spatial momentum $q\to 0$, the Green functions for the
different components coincide and were computed as \cite{Myers:2007we}
\begin{equation}
\label{eq:q0GreenFunction}
G^R = G^R_{xx} =G^R_{yy} = G^R_{zz} = 
 \frac{N_f N_c T^2}{8}\; \lim_{\rho\to\infty}\left(\rho^3 \frac{\partial_\rho E(\rho)}{E(\rho)} \right)\, ,  
\end{equation}
where the $E(\rho)$ in the denominator divides out the boundary value of the
field in the limit of large $\rho$, as discussed after
\eqref{eq:retardedThermalGreen}. The indices on the Green function denote the
components of the operators in the correlation function, in our case all
off-diagonal correlations (as~$G_{yz}$, for example) vanish.

In our case of finite baryon density, new features arise through the modified
embedding and gauge field background, which enter the equations of motion
\eqref{eq:eomFluct} for the field fluctuations. To apply the prescription to
calculate the Green function, we Fourier transform the fields as
\begin{equation}
\label{eq:fourierA}
A_\mu(\rho,\vec{x}) = \int\!\! \frac{\dd^4 k}{(2\pi)^4}\, e^{i\vec{k}\vec{x}} A_\mu(\rho,\vec{k}) \,.
\end{equation}

As above, we are free to
choose our coordinate system to give us a momentum vector of the
fluctuation with nonvanishing spatial momentum only in
$x$-direction, $\vec{k}=(\omega,q,0,0)$.

For simplicity we restrict ourselves to vanishing spatial momentum~$q=0$. In
this case the equations of motion for transversal fluctuations~$E_{y,z}$ match
those for longitudinal fluctuations~$E_x$. For a more detailed discussion see
\cite{Myers:2007we}. As an example consider the equation of motion obtained from
\eqref{eq:eomFluct} with $\sigma = 2$, determining $E_y=\omega A_2$,
\begin{equation}
\begin{split}
	\label{eq:eomEq0}
	0 =\,& E''+ \frac{\partial_\rho[\sqrt{|\det G|}G^{22}G^{44}]}{{\sqrt{|\det G|}} G^{22}G^{44}}\,
          E'- \frac{G^{00}}{G^{44}}\, \varrho_H^2 \omega^2 E \\  
	 =\,& E'' + 8 \wn^2 \frac{\tilde f}{f^2} \frac{1-\chi^2 + \rho^2 {\chi'}^2 }{\rho^4 (1-\chi^2)}\,E \\
	& + \partial_\rho \ln \left({\frac{\rho^3 f \left(1-\chi^2\right)^2}{\sqrt{1 - \chi^2 + \rho^2 {\chi'}^2 - \frac{2f(1-\chi^2)}{\tilde f^2}  (\partial_\rho \tilde A_0 )^2}}}\right) E'.
\end{split}
\end{equation}
The symbol $\wn$ denotes the dimensionless frequency $\wn=\omega/(2\pi T)$,
and we made use of the dimensionless radial coordinate
$\rho$.

In order to numerically integrate this equation, we determine local solutions of
that equation near the horizon~$\rho=1$. These can be used to compute initial
values in order to integrate~(\ref{eq:eomEq0}) forward towards the boundary. The
equation of motion~(\ref{eq:eomEq0}) has coefficients which are singular at the
horizon. According to standard methods~\cite{Bender}, the local solution of this
equation behaves as~$(\rho-\rho_{\text{H}})^\beta$, where~$\beta$ is a so-called
`index' of the differential equation. We compute the possible indices to be
\begin{equation}
\label{eq:indices}
\beta=\pm i\,\wn .
\end{equation}
Only the negative one will be retained in the following, since it casts the
solutions into the physically relevant incoming waves at the horizon and
therefore satisfies the incoming wave boundary condition. The solution $E$ can
be split into two factors, which are $(\rho-1)^{-i\wn}$ and some function
$F(\rho)$, which is regular at the horizon. The first coefficients of a series
expansion of $F(\rho)$ can be found recursively as described
in~\cite{Teaney:2006nc,Kovtun:2006pf}. At the horizon the local solution then
reads
\begin{equation}
\begin{aligned}
\label{eq:localSolutions}
E(\rho)  =\; & (\rho-1)^{-i\wn}\, F(\rho)\\
          =\; & (\rho-1)^{-i\wn} \left [1+\frac{i\wn}{2}(\rho-1)+
\cdots \right ].
\end{aligned}
\end{equation}
So, $F(\rho)$ asymptotically assumes values
\begin{equation}
\label{eq:startingValues}
F(\rho=1)=1,\qquad \partial_\rho F(\rho)\Big|_{\rho=1}=\frac{i\wn}{2}\, .
\end{equation}

For the calculation of numbers, we have to specify the baryon density~$\tilde d$
and the mass parameter $\chi_0\sim m_q/T$ to obtain the embeddings $\chi$ used
in \eqref{eq:eomEq0}. Then we obtain a solution for a given frequency $\wn$
using initial values (\ref{eq:localSolutions}) and (\ref{eq:startingValues}) in
the equation of motion~(\ref{eq:eomEq0}). This eventually gives us the numerical
solutions for $E(\rho)$.

Spectral functions are then obtained by combining (\ref{eq:q0GreenFunction}) and 
\eqref{eq:specDensity},
\begin{equation}
        \R(\omega,0) = - \frac{N_f N_c T^2}{4}\; \mathrm{Im}\lim_{\rho\to\infty}\left(\rho^3 \frac{\partial_\rho E(\rho)}{E(\rho)} \right).
\end{equation}

\subsubsection{Results for spectral functions}
\label{sec:vectorResults}

\begin{figure}
		\psfrag{w}{$\wn$}
		\psfrag{RmR0}{$\mathfrak{R}(\wn,0)-\mathfrak{R}_0$}
		\psfrag{dt025}{\footnotesize $\tilde d = 0.25$}
		\psfrag{chi001}{\footnotesize $\chi_0 = 0.1$}
		\psfrag{chi005}{\footnotesize $\chi_0 = 0.5$}
		\psfrag{chi007}{\footnotesize $\chi_0 = 0.7$}
		\psfrag{chi008}{\footnotesize $\chi_0 = 0.8$}
        \includegraphics[width=\linewidth]{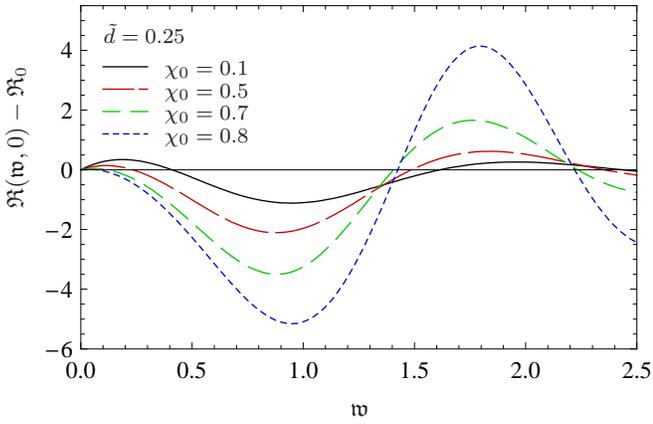}
        \caption{The finite temperature part of the spectral function~$\R-\R_0$
			(in units of~$N_f N_c T^2/4$) at finite baryon density~$\tilde d$.
			The maximum grows and shifts to smaller frequencies
			as~$\chi_0$ is increased towards~$\chi_0=0.7$, but then turns around
			to approach larger frequency values.}
        \label{fig:vectorPeaksDt025light}
\end{figure}

\begin{figure}
		\psfrag{w}{$\wn$}
		\psfrag{RmR0}{$\mathfrak{R}(\wn,0)-\mathfrak{R}_0$}
		\psfrag{dt025}{\footnotesize $\tilde d = 0.25$}
		\psfrag{chi008}{\footnotesize $\chi_0 = 0.8$}
		\psfrag{chi0094}{\footnotesize $\chi_0 = 0.94$}
		\psfrag{chi00962}{\footnotesize $\chi_0 = 0.962$}
        \includegraphics[width=\linewidth]{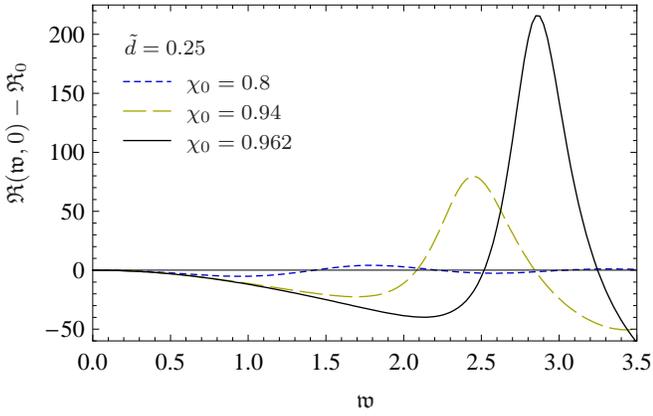}
        \caption{The finite temperature part of the spectral function~$\R-\R_0$
			(in units of~$N_f N_c T^2/4$) at finite baryon density~$\tilde d$.
			In the regime of~$\chi_0$ shown here, the peak shifts to larger
			frequency values with increasing $\chi_0$.}
        \label{fig:vectorPeaksDt025heavy}
\end{figure}

\begin{figure}
		\psfrag{w}{$\wn$}
		\psfrag{RmR0}{$\mathfrak{R}(\wn,0)-\mathfrak{R}_0$}
		\psfrag{dt025}{\footnotesize $\tilde d = 0.25$}
		\psfrag{chi0962}{\footnotesize $\chi_0 = 0.962$}
		\psfrag{n0}{\footnotesize $n=0$}
		\psfrag{n1}{\footnotesize $n=1$}
		\psfrag{n2}{\footnotesize $n=2$}
		\psfrag{n3}{\footnotesize $n=3$}
		\includegraphics[width=\linewidth]{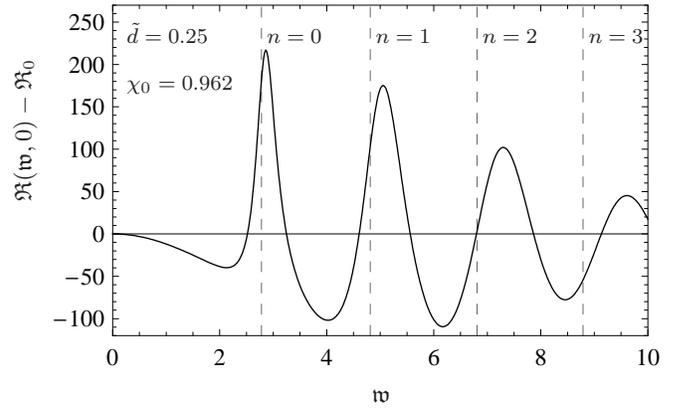}
		\caption{The finite temperature part~$\R-\R_0$ of the spectral function
			(in units of~$N_f N_c T^2/4$) at finite baryon density~$\tilde d$.
			The oscillation peaks narrow and get more pronounced compared to
			smaller~$\chi_0$. Dashed vertical lines show the meson mass spectrum
			given by equation~\eqref{eq:massFormula2}.}
        \label{fig:lineSpectrumDt025light}
\end{figure}

\begin{figure}
		\psfrag{w}{$\wn$}
		\psfrag{R}{$\mathfrak{R}(\wn,0)$}
		\psfrag{dt025}{\footnotesize $\tilde d = 0.25$}
		\psfrag{chi0999}{\footnotesize $\chi_0 = 0.999$}
		\psfrag{n0}{\footnotesize $n=0$}
		\psfrag{n1}{\footnotesize $n=1$}
		\psfrag{n2}{\footnotesize $n=2$}
		\psfrag{n3}{\footnotesize $n=3$}
		\includegraphics[width=\linewidth]{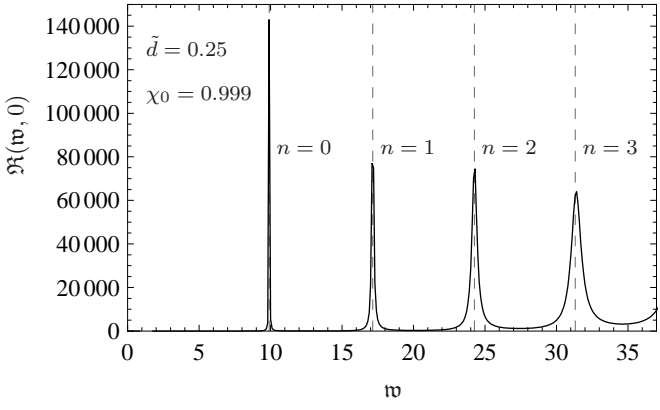}
        \caption{The spectral function $\mathfrak{R}$ (in units of~$N_f N_c
			T^2/4$) at finite baryon density~$\tilde d$. At large $\chi_0$, as
			here, the peaks approach the dashed drawn line spectrum given
			by~\eqref{eq:massFormula2}.}
        \label{fig:lineSpectrumDt025heavy}
\end{figure}

\begin{figure}
		\psfrag{w}{$\wn$}
		\psfrag{R}{$\mathfrak{R}$}
		\psfrag{R0}{$\mathfrak{R}_0$}
		\psfrag{Rtemp}{$\mathfrak{R}(\wn,0)$}
		\psfrag{dt025}{\footnotesize $\tilde d = 0.25$}
		\psfrag{chi00997}{\footnotesize $\chi_0 = 0.997$}
		\includegraphics[width=\linewidth]{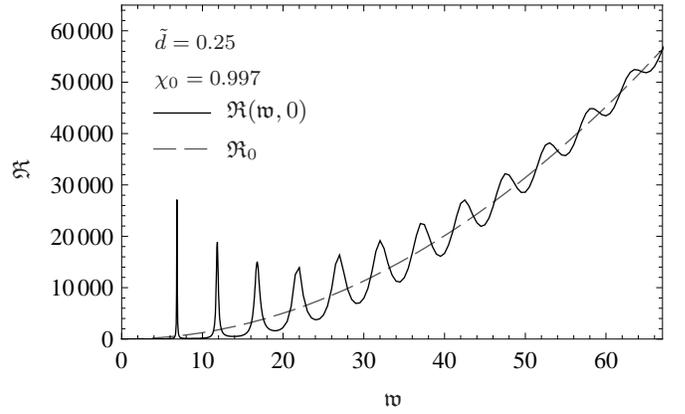}
        \caption{The thermal spectral function $\mathfrak{R}$ (in units
			of~$N_f N_c T^2/4$) compared to the zero temperature result
			$\mathfrak{R}_0$.}
		\label{fig:specTempZeroTemp}
\end{figure}

We now discuss the resulting spectral functions at finite baryon density, and
observe crucial qualitative differences compared to the case of vanishing baryon
density. In figures~\ref{fig:vectorPeaksDt025light}
to~\ref{fig:lineSpectrumDt025heavy}, some examples for the spectral function at
fixed baryon density $n_B\propto \tilde d$ are shown. To emphasize the resonance
peaks, in some plots we subtract the quantity
\begin{equation}
	\mathfrak{R}_0 = N_f N_c T^2\, \pi\wn^2,
\end{equation}
around which the spectral functions oscillate,
cf.~figure~\ref{fig:specTempZeroTemp}.

The graphs are obtained for a value of $\tilde d$ above $\tilde d^*$~(given by
\eqref{eq:dcrit}), where the fundamental phase transition does not occur. The
different curves in these plots show the spectral functions for different quark
masses, corresponding to different positions on the solid blue line in the phase
diagram shown in figure~\ref{fig:phaseDiagram}. Regardless whether we chose
$\tilde d$ to be below or above the critical value of $\tilde d$, we observe the
following behavior of the spectral functions with respect to changes in the
quark mass to temperature ratio.

Increasing the quark mass from zero to small finite values results in more and
more pronounced peaks of the spectral functions. This eventually leads to the
formation of resonance peaks in the spectrum. At small masses, though, there are
no narrow peaks. Only some maxima in the spectral functions are visible. At
the same time as these maxima evolve into resonances with increasing quark mass,
their position changes and moves to lower freqencies $\wn$, see
figure~\ref{fig:vectorPeaksDt025light}. This behavior was also observed for the
case of vanishing baryon density in ~\cite{Myers:2007we}.

However, further increasing the quark mass leads to a crucial difference to the
case of vanishing baryon density. Above a value $m^{\text{turn}}$ of the quark
mass, parametrized by $\chi_0^{\text{turn}}$, the peaks change their direction
of motion and move to larger values of $\wn$, see
figure~\ref{fig:vectorPeaksDt025heavy}. Still the maxima evolve into more and
more distinct peaks.

Eventually at very large quark masses, given by $\chi$ closer and closer to 1,
the positions of the peaks 
asymptotically reach exactly those frequencies which 
correspond to the masses of the vector mesons at zero temperature
\cite{Kruczenski:2003be}. In our coordinates, these masses are given by

\begin{equation}
\label{eq:massFormula2}
M= \frac{L_\infty}{R^2}\,\sqrt{2 (n+1)(n+2)}\, ,
\end{equation}
where $n$ labels the Kaluza-Klein modes arising from the D7-brane wrapping
$S^3$, and~$L_\infty$ is the radial distance in~$(8\mathord{,}9)$-direction
between the stack of D3-branes and the D7, evaluated at the $AdS$-boundary,
\begin{equation}
\label{eq:L}
L_\infty=\lim_{\varrho\to\infty} \varrho\chi(\varrho)  \, .
\end{equation}

The formation of a line-like spectrum can be interpreted as the evolution of
highly unstable quasi-particle excitations in the plasma into quark bound
states, finally turning into nearly stable vector mesons,
cf.~figures~\ref{fig:lineSpectrumDt025light}
and~\ref{fig:lineSpectrumDt025heavy}.

\medskip 

We now consider the turning behavior of the resonance peaks shown in
figures~\ref{fig:vectorPeaksDt025light} and~\ref{fig:vectorPeaksDt025heavy}.
There are two different scenarios, depending on whether the quark mass is small
or large.

First, when the quark mass is very small~$m_q\ll T$, we are in the regime of the
phase diagram corresponding to the right half  of
figure~\ref{fig:phaseDiagram}. In this regime the influence of the Minkowski
phase is negligible, as we are deeply inside the black hole phase. We therefore
observe only broad structures in the spectral functions, instead of peaks.

Second, when the quark mass is very large, $m_q\gg T$, or equivalently the
temperature is very small, the quarks behave just as they would at zero
temperature, forming a line-like spectrum. This regime corresponds to the left
side of the phase diagram in figure~\ref{fig:phaseDiagram}, where all curves of
constant $\tilde d$ asymptote to the Minkowski phase.

The turning of the resonance peaks is associated to being in the 
first or in the second regime. At~$\chi_0^{\text{turn}}$ the two regimes
are connected and none of them is dominant. 

The turning behavior is best understood by following a line of constant density
$\tilde d$ in the phase diagram of figure~\ref{fig:phaseDiagram}. Consider for
instance the solid blue line in figure~\ref{fig:phaseDiagram}, starting at large
temperatures/small masses on the right of the plot. First, we are deep in the
unshaded region ($n_B\not =0$), far inside the black hole phase. Moving along to
lower $T/\bar M$, the solid blue line in figure~\ref{fig:phaseDiagram} rapidly
bends upwards, and asymptotes to both the line corresponding to the onset of the
fundamental phase transition, as well as to the separation line between black
hole and Minkowski phase (gray region).

This may be interpreted as the quarks joining in bound states. Increasing the
mass further, quarks form almost stable mesons, which give rise to resonance
peaks at larger frequency if the quark mass is increased. The confined and
deconfined phase are coexistent asymptotically for $T/\bar M \to 0$.

We also observe a dependence of $\chi_0^{\text{turn}}$ on the baryon density. As
the baryon density is increased from zero, the value of $\chi_0^{\text{turn}}$
decreases.

Figures~\ref{fig:lineSpectrumDt025heavy} and~\ref{fig:specTempZeroTemp} show
that higher~$n$ excitations from the Kaluza-Klein tower are less stable. While
the first resonance peaks in this plot are very narrow, the following peaks show
a broadening with decreasing amplitude.

\begin{figure}
	\includegraphics[width=.48\linewidth]{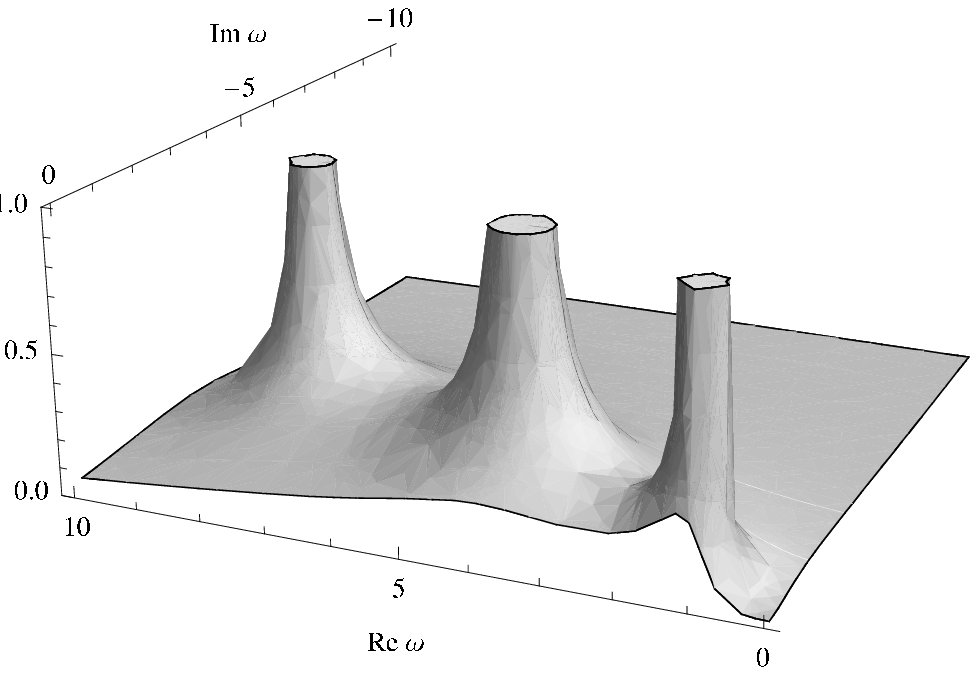}
	\hfill
	\includegraphics[width=.48\linewidth]{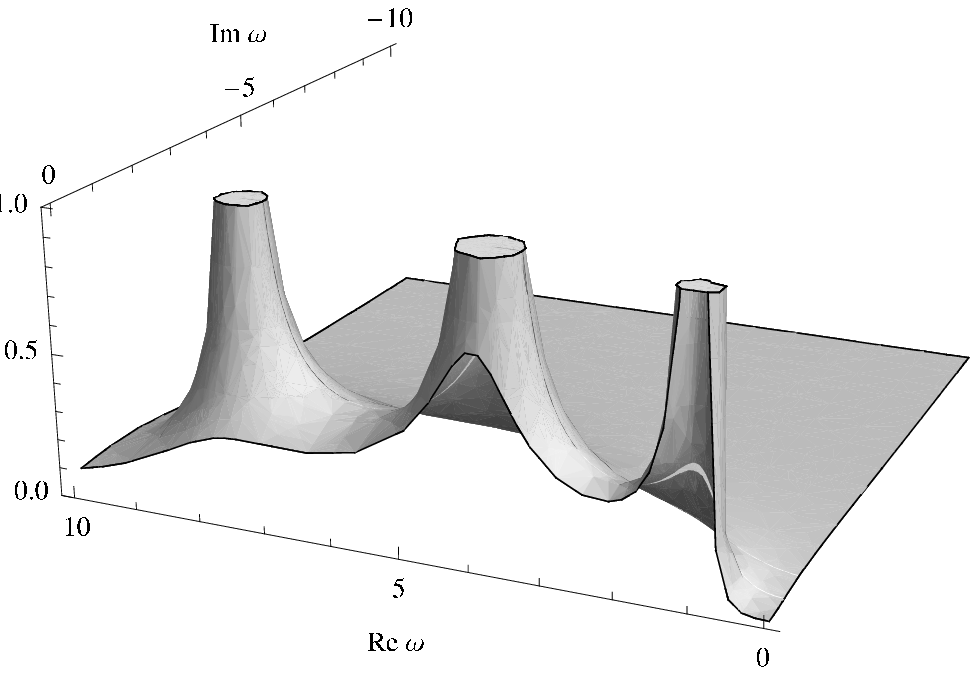}
	\caption{Qualitative relation between the location of the poles in the
		complex frequency plane and the shape of the spectral functions on the
		real $\omega$ axis. The function plotted here is an example for the
		imaginary part of a correlator. Its value on the real $\omega$ axis
		represents the spectral function. The poles in the right plot are closer
		to the real axis and therefore there is more structure in the spectral
		function.}
	\label{fig:polesExample}
\end{figure}

This broadening of the resonances is due to the behaviour of the quasinormal
modes of the fluctuations, which correspond to the poles of the correlators in
the complex~$\omega$ plane, as described in the example
\eqref{eq:quasiNormalModes} and sketched in figure~\ref{fig:polesExample}. The
location of the resonance peaks on the real frequency axis corresponds to the
real part of the quasinormal modes. It is a known fact that the the quasinormal
modes develop a larger real and \emph{imaginary} part at higher $n$. So the
sharp resonances at low~$\wn$, which correspond to quasiparticles of long
lifetime, originate from poles whith small imaginary part. For higher
excitations in $n$ at larger~$\wn$, the resonances broaden and get damped due to
larger imaginary parts of the corresponding quasi normal modes.

For increasing mass we described above that the peaks of the spectral functions
first move to smaller frequencies until they reach the turning point
$m^{\text{turn}}$. Further increasing the mass leads to the peaks moving to
larger frequencies, asymptotically approaching the line spectrum. This behavior
can be translated into a movement of the quasinormal modes in the complex plane.
It would be interesting to compare our results to a direct calculation of the
quasinormal modes of vector fluctuations in
analogy to~\cite{Hoyos:2006gb}.

In~\cite{Hoyos:2006gb} the quasinormal modes 
are considered for scalar fluctuations exclusively, at vanishing
baryon density. The authors observe that starting 
from the massless case, the real part of the quasinormal frequencies
increases with the quark mass first, and then turns around to decrease.   
This behavior agrees with 
the peak movement for scalar spectral functions observed in~\cite[figure 9]{Myers:2007we}
(above the fundamental phase transition,~$\chi_0\le 0.94$) 
where the scalar meson resonances move to 
higher frequency first, turn around and
move to smaller frequency increasing the mass further. These results
do not contradict the present work since we consider vector modes exclusively.
The vector meson spectra considered in~\cite{Myers:2007we} at vanishing baryon density   
only show peaks moving to smaller frequency as the quark mass is increased.  
Note that the authors there continue to consider black hole embeddings below the 
fundamental phase transition which are only metastable, the Minkowski 
embeddings being thermodynamically favored. At small baryon density
and small quark mass our spectra are virtually coincident with those of~\cite{Myers:2007we}.  
In our case, at finite baryon density, black hole embeddings are favored
for all values of the mass over temperature ratio. At small values
of~$T/\bar M$ in the phase diagram of figure~\ref{fig:phaseDiagram}, 
we are very close to the Minkowski regime, temperature effects
are small, and the meson mass is proportional to the quark
mass as in the supersymmetric case. Therefore, the peaks 
in the spectral function move to the right~(higher frequencies) as function
of increasing quark mass.

The turning point in the location of the peaks is a consequence 
of the transition between two regimes, i.e.\ the temperature-dominated 
one also observed in~\cite{Myers:2007we}, and the potential-dominated  
one which asymptotes to the supersymmetric spectrum.

We expect the physical interpretation of the left-moving of the 
peaks in the temperature-dominated regime to be related to the
strong dissipative effects present in this case. This is consistent 
with the large baryon diffusion coefficient present in this regime 
as discussed in section~\ref{sec:baryonDiff} and shown  
in figure~\ref{fig:diffusionConstants}. A detailed 
understanding of the physical picture in this regime requires a 
quantitative study of the quasipaticle behavior which we 
leave to future work.

In our approach it is straightforward to investigate 
the $T\to 0$ limit since black hole embeddings are 
thermodynamically favored even near~$T=0$ at finite baryon density. 
We expect that a right-moving of the peaks consistent with
the SUSY spectrum should also be observable for Minkowski embeddings 
at vanishing baryon density. However this has not been 
investigated for vector modes neither in~\cite{Hoyos:2006gb} 
nor in~\cite{Myers:2007we}.

\section[Spectral functions at finite isospin density]{ 
Spectral functions at finite isospin density}
\label{sec:isospinMu}

\subsection{Radially varying $SU(2)$-background gauge field}
\label{sec:varyChemPot}
In order to examine the case $N_f=2$ in the strongly coupled plasma, we extend
our previous analysis of vector meson spectral functions to a chemical potential
with $SU(2)$-flavor (isospin) structure. Starting from the general action
\begin{equation}
\label{eq:isoAction}
S_{\text{iso}}=-T_r T_{D7} \int\!\!\dd^8\xi\;\sqrt{|\det(g+\hat F)|} \, ,   
\end{equation} 
we now consider field strength tensors
\begin{equation}
\label{eq:isoF}
\hat F_{\mu\nu}=\sigma^a\,\left (2\partial_{[\mu}\hat A^a_{\nu]}
  +\frac{\varrho_H^2}{2\pi\alpha'}f^{abc}\hat A_\mu^b \hat A_\nu^c\right)\, ,
\end{equation}
with the Pauli matrices~$\sigma^a$ and $\hat A$ given by 
equation~\eqref{eq:hatA}. The factor~$\varrho_H^2/(2\pi\alpha')$ is due
to the introduction of dimensionless fields as described below~\eqref{eq:chemPotLimit}. 
In order to obtain a finite isospin-charge
density~$n_I$ and its conjugate chemical potential~$\mu_I$, we introduce an
$SU(2)$-background gauge field~$\tilde A$~\cite{Erdmenger:2007ap}  
\begin{equation}
\label{eq:isoBackgrd}
\tilde A^3_{0}\sigma^3= \tilde A_0(\rho) 
\left (\begin{array}{c c}
1 & 0\\ 
0& -1
\end{array}\right ) \, .
\end{equation}
This specific choice of the 3-direction in flavor space as well as  spacetime
dependence simplifies the isospin background field strength, such that we get
two copies of the baryonic background~$\tilde F_{\rho 0}$ on the diagonal of the
flavor matrix,
\begin{equation}
\tilde F_{\rho 0}\,\sigma^3=
\left (\begin{array}{c c}
\partial_ \rho \tilde A_0& 0\\ 
0& -\partial_ \rho \tilde A_0
\end{array}\right )\,  .  
\end{equation}
The action for the isospin  
background differs from the action~\eqref{eq:actionEmbeddingsAt} for the baryonic
background only by a group theoretical factor:  
The factor~$T_r=1/2$~(compare  
\eqref{eq:isoAction}) replaces the
baryonic factor~$N_f$ in equation~(\ref{eq:dbiAction}), 
which arises by summation over the
$U(1)$ representations. We can thus use the
embeddings~$\chi(\rho)$ and background field solutions~$\tilde A_0(\rho)$ of  
the baryonic case of~\cite{Kobayashi:2006sb}, listed here in
section~\ref{sec:backAndBranes}. As before, we collect the induced metric~$g$
and the background field strength~$\tilde F$ in the background tensor
$G=g+\tilde F$.

We apply the background field method in analogy to the baryonic case examined in
section~\ref{sec:baryonMu}. As before, we obtain the quadratic action by
expanding the determinant and square root in fluctuations $A^a_\mu$.  The term
linear in fluctuations again vanishes by the equation of motion for
our background field. This leaves the quadratic action
\begin{multline}
\label{eq:quadIsoAction}
S^{(2)}_{\text{iso}} = {\varrho_H}  (2\pi^2 R^3) T_{D7} T_r
\int\limits_1^\infty\!\! \dd\rho\,\dd^4x\; \sqrt{\left|\det G\right|}\\   
  {\times}\Big[G^{\mu\mu'} G^{\nu\nu'} \Big (
       \partial_{[\mu}A^a_{\nu]}\partial_{[\mu'}A^a_{\nu']}
		\hphantom{G^{\mu\mu'}G^{\mu\mu'} G^{\mu\mu'} }\\ 
       \hphantom{G^{\mu\mu'} }+\frac{{\varrho_H}^4}{(2\pi\alpha')^2}(\tilde A_0^3)^2 f^{ab3} f^{ab'3} A_{[\mu}^b\delta_{\nu]0}
        A_{[\mu'}^{b'}\delta_{\nu']0} \Big)\\  
       +(G^{\mu\mu'} G^{\nu\nu'}\!\!\!\! -G^{\mu'\mu} G^{\nu'\nu})
       \frac{{\varrho_H}^2}{2\pi\alpha'} \tilde A_0^3 f^{ab3} \partial_{[\mu'}A^a_{\nu']} A_{[\mu}^{b}\delta_{\nu]0} \Big] .
\end{multline}
Note that besides the familiar Maxwell term, two other terms appear, which are 
due to the non-Abelian structure. One of the new terms  
depends linearly, the other quadratically on the background 
gauge field~$\tilde A$ and both contribute
nontrivially to the dynamics.
The equation of motion for gauge field fluctuations on the D7-brane is
\begin{align}
\label{eq:eomIsoFluct}  
0=\;&\partial_\kappa \left[ \sqrt{\left|\det G\right|}
 \left( G^{\nu\kappa} G^{\sigma\mu} - G^{\nu\sigma} G^{\kappa\mu} \right)
 \check F_{\mu\nu}^a \right] \\ \nonumber   
 & - \sqrt{\left|\det G\right|}
  \frac{{\varrho_H}^2}{2\pi\alpha'} \tilde A_0^3 f^{ab3} \left( G^{\nu 0} G^{\sigma\mu} 
  - G^{\nu\sigma} G^{0\mu} \right)\check F_{\mu\nu}^b
   \, ,  
\end{align}
with the modified field strength linear in fluctuations 
$\check{F}^a_{\mu\nu}=2\partial_{[\mu}A^a_{\nu]}+f^{ab3}
   \tilde A_0^3( \delta_{0\mu} A_\nu^b+ \delta_{0\nu} A_\mu^b){\varrho_H^2}/{(2\pi\alpha')}$.

Integration by parts of~\eqref{eq:quadIsoAction} and application of~\eqref{eq:eomIsoFluct} 
yields the on-shell action  
\begin{align}
\label{eq:onShellAction}
S^{\text{on-shell}}_{\text{iso}}=\; 
 &\varrho_H T_r T_{D7}  \pi^2 R^3 \int\!\! \dd^4x\, \sqrt{\left|\det G\right|}\nonumber \\
 & \times\left. \left (  
 G^{\nu 4} G^{\nu'\mu}- G^{\nu\nu'} G^{4\mu}
  \right ) A_{\nu'}^a \check F^a_{\mu\nu}\right |_{\rho_H}^{\rho_B}\, . 
\end{align}
The three flavor field equations of motion~(flavor index $a=1,2,3$) for fluctuations in 
transversal Lorentz-directions $\alpha=2,3$ can again be written in terms of the 
combination~$E^a_T=q A^a_0+\omega A^a_\alpha$.
At vanishing spatial momentum~$q=0$ we get
\begin{align}
\label{eq:eomAalphaSplitFlav1}
0 =\;& {E_T^1}''+\frac{\partial_\rho(\sqrt{\left|\det G\right|} G^{44} G^{22})}{\sqrt{\left|\det G\right|} G^{44} G^{22}} {E_T^1}' \\
     & \hphantom{{E_T^1}''}-\frac{G^{00}}{G^{44}}\left[(\varrho_H \omega)^2+\left({\frac{\varrho_H^2}{2\pi\alpha'}}\tilde A^3_0\right)^2\right]E^1_T \nonumber\\
     & \hphantom{{E_T^1}''}+\frac{2 i \varrho_H \omega G^{00}}{G^{44}} {\frac{\varrho_H^2}{2\pi\alpha'}}\tilde A^3_0 E^2_T\, ,\nonumber \\[\medskipamount]
\label{eq:eomAalphaSplitFlav2}
0 =\;& {E_T^2}'' +\frac{\partial_\rho(\sqrt{\left|\det G\right|} G^{44} G^{22})}{\sqrt{\left|\det G\right|} G^{44} G^{22}} {E_T^2}' \\
     & \hphantom{{E_T^2}''}-\frac{G^{00}}{G^{44}}\left[(\varrho_H \omega)^2+\left({\frac{\varrho_H^2}{2\pi\alpha'}}\tilde A^3_0\right)^2\right]E^2_T \nonumber \\
     & \hphantom{{E_T^2}''}- \frac{2 i \varrho_H \omega G^{00}}{G^{44}} {\frac{\varrho_H^2}{2\pi\alpha'}}\tilde A^3_0 E^1_T \, ,\nonumber \\[\medskipamount]
\label{eq:eomAalphaSplitFlav3}
\hphantom{.} 0 =\;&  {E_T^3}'' +\frac{\partial_\rho(\sqrt{\left|\det G\right|} G^{44} G^{22})}{\sqrt{\left|\det G\right|} G^{44} G^{22}} {E_T^3}' - \frac{G^{00} (\varrho_H\omega)^2}{G^{44}}E^3_T  \, .
\end{align}
Note that we use the dimensionless background gauge 
field~$\tilde A_0^3=\bar A_0^3 (2\pi \alpha')/\varrho_H$ and $\varrho_H=\pi T R^2$.
Despite the presence of the new non-Abelian terms, 
at vanishing spatial momentum the equations of motion for 
longitudinal fluctuations are the same as the transversal equations~\eqref{eq:eomAalphaSplitFlav1},
\eqref{eq:eomAalphaSplitFlav2} and \eqref{eq:eomAalphaSplitFlav3},  
such that~$E=E_T=E_L$.

Note at this point that there are two essential differences which distinguish this setup from the approach with a constant
potential~$\bar A_0^3$ at vanishing mass followed in~\cite{Erdmenger:2007ap}. First, the inverse metric coefficients $g^{\mu\nu}$ contain the 
embedding function $\chi(\rho)$ computed with varying background gauge field. Second,  
the background gauge field~$\bar A_0^3$ giving rise to the chemical potential now depends on $\rho$. 

Two of the ordinary second order differential equations~(\ref{eq:eomAalphaSplitFlav1}), 
(\ref{eq:eomAalphaSplitFlav2}), (\ref{eq:eomAalphaSplitFlav3}) are coupled through their flavor structure. Decoupling can
be achieved by transformation to the flavor combinations~\cite{Erdmenger:2007ap}
\begin{equation}
\label{eq:flavorTrafo}
X =E^1+ i E ^2,\;\; \; \;     
Y =E^1- i E ^2\, .
\end{equation}
The equations of motion for these fields are given by
\begin{align}
\label{eq:eomX}
0=&\;{X}'' +\frac{\partial_\rho(\sqrt{\left| \det G \right|} G^{44} G^{22})}{\sqrt{\left|\det G\right|} G^{44} G^{22}} {X}' \\
 \nonumber
 & \;\hphantom{{X}''}- 4 {\frac{\varrho_H^4}{R^4}}\frac{G^{00}}{G^{44}}\left(\wn- \mathfrak{m}\right)^2 X\, , \\[\medskipamount]    
\label{eq:eomY}
0=&\;{Y}'' +\frac{\partial_\rho(\sqrt{\left|\det G\right|} G^{44} G^{22})}{\sqrt{\left|\det G\right|} G^{44} G^{22}} {Y}' \\
 \nonumber
 & \;\hphantom{{X}''}- 4 {\frac{\varrho_H^4}{R^4}}\frac{G^{00}}{G^{44}}\left(\wn+ \mathfrak m\right)^2 Y\, , \\[\medskipamount]
\label{eq:eomE3}
0=&\;{E^3}'' +\frac{\partial_\rho(\sqrt{\left|\det G\right|} G^{44} G^{22})}{\sqrt{\left|\det G\right|} G^{44} G^{22}} {E^3}' -
4 \frac{\varrho_H^4}{R^4}\frac{G^{00}}{G^{44}} \wn^2E^3\, , 
\end{align}
with dimensionless~$\mathfrak m= \bar A^3_0 /(2\pi T)$ and $\wn=\omega/(2\pi T)$.
Proceeding as described in section~\ref{sec:baryonMu}, we determine the local
solution of~(\ref{eq:eomX}),~(\ref{eq:eomY}) and~(\ref{eq:eomE3}) at the
horizon. The indices turn out to be
\begin{equation}
\label{eq:isospinIndices}
\beta = \pm i \left[\wn \mp\frac{\bar A^3_0(\rho=1)}{(2\pi T)}\right ]\, .
\end{equation}
Since $\bar A^3_0(\rho=1)=0$ in the setup considered here, we are left with the
same index as in \eqref{eq:indices} for the baryon case. Therefore, here the
chemical potential does not influence the singular behavior of the fluctuations
at the horizon. The local solution coincides to linear order with the baryonic
solution given in~(\ref{eq:localSolutions}).

Application of the recipe described in section~\ref{sec:specFuncs} yields the
spectral functions of flavor current correlators shown in
figures~\ref{fig:isoLinesDt025} and~\ref{fig:isoXYsplitDt025}. Note that after
transforming to flavor combinations~$X$ and~$Y$, given in
\eqref{eq:flavorTrafo}, the diagonal elements of the propagation submatrix in
flavor-transverse~$X,\, Y$ directions vanish, $G_{XX}=G_{YY}=0$, while the
off-diagonal elements give non-vanishing contributions.
However, the component $E^3$, longitudinal in flavor space, is
not influenced by the isospin chemical potential, such that~$G_{E^3E^3}$ is
nonzero, while other combinations with $E^3$ vanish~(see~\cite{Erdmenger:2007ap}
for details).

Introducing the chemical potential as described above for a zero-temperature
$AdS_5\times S^5$ background, we obtain the gauge field correlators in analogy
to~\cite{Freedman:1998tz}. The resulting spectral function for the field theory
at zero temperature but finite chemical potential and density
$\mathfrak{R}_{0,\mathrm{iso}}$ is given by
\begin{equation}
\label{eq:zeroTspecFuncWithIsospin}
\mathfrak{R}_{0,\mathrm{iso}}=\frac{N_c T^2 T_r}{4}4\pi(\wn\pm \mn_\infty)^2   \, ,  
\end{equation}
with the dimensionless chemical 
potential~$\mn_\infty=\lim_{\rho\to\infty}\bar A_0^3/(2\pi T)=\mu/(2\pi T)$.
Note that~\eqref{eq:zeroTspecFuncWithIsospin} is independent of the temperature.
This part is always subtracted when we consider spectral functions at finite
temperature, in order to determine the effect of finite temperature separately,
as we did in the baryonic case.

\subsection{Results 
at finite isospin density}
\begin{figure}
	\psfrag{0}[cc]{$0$}
	\psfrag{0.5}[cc]{$0.5$}
	\psfrag{1.5}[cc]{$1.5$}
	\psfrag{2.5}[cc]{$2.5$}
	\psfrag{1}[cc]{$1$}
	\psfrag{2}[cc]{$2$}
	\psfrag{-2}[cc]{$-2$}
	\psfrag{-1}[cc]{$-1$}	
	\psfrag{YX}{$XY$}
	\psfrag{XY}{ $YX$}
	\psfrag{R}[bc]{$\mathfrak{R}-\mathfrak{R}_0$}
	\psfrag{w}{$\wn$}
        \includegraphics[width=.9\linewidth]{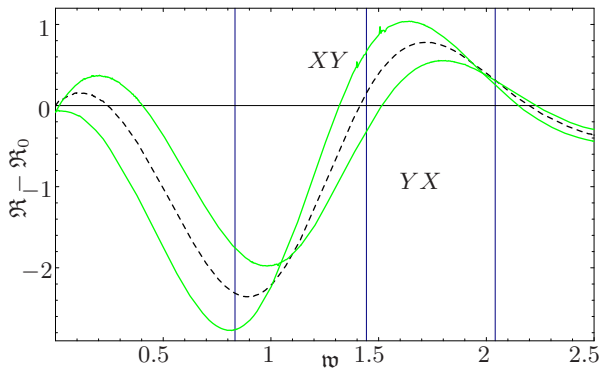}
        \caption{
        The finite temperature part of spectral
        functions~$\R_{\mathrm{iso}}-\mathfrak{R}_{0,\mathrm{iso}}$ (in units
        of~$N_c T^2 T_r/4$) of currents dual to fields~$X,\,Y$ 
        are shown versus~$\wn$. The dashed line shows the baryonic
        chemical potential case, the solid curves show the spectral functions in
        presence of an isospin chemical potential.
        Plots are generated for~$\chi_0=0.5$ and~$\tilde d=0.25$. 
        The combinations $X Y$ and $YX$ split in opposite directions 
        from the baryonic spectral function.
                       }
        \label{fig:isoLinesDt025}
\end{figure}

\begin{figure}
	\psfrag{0}[cc]{$0$}
	\psfrag{1}[cc]{$1$}
	\psfrag{2}[cc]{$2$}
	\psfrag{4}[cc]{$4$}
	\psfrag{6}[cc]{$6$}
	\psfrag{8}[cc]{$8$}
	\psfrag{10}[cc]{$10$}
	\psfrag{1000}[cc]{$1000$}
	\psfrag{2000}[cc]{$2000$}
	\psfrag{3000}[cc]{$3000$}
	\psfrag{4000}[cc]{$4000$}	
	\psfrag{XY}{\tiny $XY$}
	\psfrag{YX}{\tiny $YX$}
	\psfrag{E3E3}{\tiny $E_3 E_3$}
	\psfrag{nis0}{\tiny $n=0$}
	\psfrag{nis1}[cc]{\tiny $n=1$}
	\psfrag{R}[ct]{$\mathfrak{R}-\mathfrak{R}_0$}
	\psfrag{-}{}
	\psfrag{R0}{}
	\psfrag{w}{$\wn$}
	\includegraphics[width=.9\linewidth]{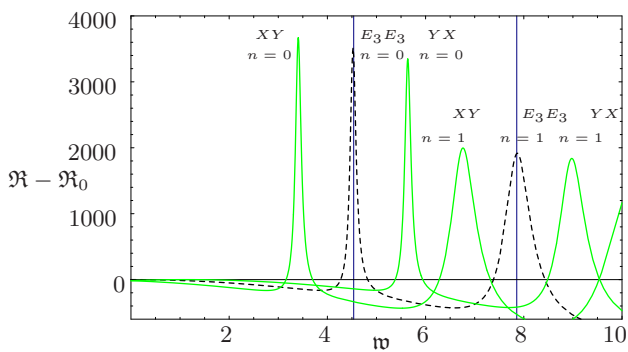}
        \caption{
        A comparison between the finite temperature part of the 
        spectral functions~$\mathfrak{R}_{XY}$ and $\mathfrak{R}_{YX}$~(solid lines)
         in the two flavor directions~$X$ 
        and~$Y$ transversal to the chemical potential
        is shown in units of~$N_c T^2 T_r/4$ for large quark mass to temperature
        ratio~$\chi_0=0.99$ and~$\tilde d=0.25$. The spectral 
        function~$\mathfrak{R}_{E^3E^3}$ along the~$a=3$-flavor direction is shown as 
        a dashed line. We observe a splitting of the line expected at the lowest 
        meson mass at~$\wn=4.5360$~($n=0$). The resonance is shifted to lower
		frequencies for~$\mathfrak{R}_{XY}$ 
        and to higher ones for~$\mathfrak{R}_{YX}$, while it remains in place
        for ~$\mathfrak{R}_{E^3E^3}$. The second meson resonance peak~($n=1$)
        shows a similar behavior. So the different flavor combinations
        propagate differently and have distinct quasiparticle resonances.
          }
        \label{fig:isoXYsplitDt025}
\end{figure}

In figure~\ref{fig:isoLinesDt025} we compare typical spectral functions found
for the isospin case~(solid lines) with that found in the baryonic
case~(dashed line). While the qualitative behavior of the isospin spectral
functions agrees with the one of the baryonic spectral functions, there
nevertheless is a quantitative difference for the components $X,\,Y$, which are
transversal to the background in flavor space. We find that the propagator for
flavor combinations~$G_{YX}$ exhibits a spectral function for which the zeroes
as well as the peaks are shifted to higher frequencies, compared to the Abelian
case curve. For the spectral function computed from~$G_{XY}$, the opposite is
true. Its zeroes and peaks appear at lower frequencies. As seen from
figure~\ref{fig:isoXYsplitDt025}, also the quasiparticle resonances of these two
different flavor correlations show distinct behavior. The quasiparticle
resonance peak in the spectral function~$\R_{YX}$ appears at higher frequencies
than expected from the vector meson mass formula~$\eqref{eq:massFormula}$~(shown
as dashed grey vertical lines in figure~\ref{fig:isoXYsplitDt025}). The other
flavor-transversal spectral function~$\R_{XY}$ displays a resonance at lower
frequency than observed in the baryonic curve. The spectral function for the
third flavor direction~$\R_{E^3E^3}$ behaves as~$E$ in the baryonic case.

This may be viewed as a splitting of the resonance peak into three distinct
peaks with equal amplitudes. This is due to the fact that we explicitly break
the symmetry in flavor space by our choice of the background field~$\tilde
A^3_0$. Decreasing the chemical potential reduces the distance of the two outer
resonance peaks from the one in the middle and therefore the splitting is
reduced.

The described behavior resembles the mass splitting of mesons in presence of a
isospin chemical potential expected to occur in
QCD~\cite{He:2005nk,Chang:2007sr}. A linear dependence of the separation
of the peaks on the chemical potential is expected.
Our observations confirm this behavior. Since our vector mesons are
isospin triplets and we break the isospin symmetry explicitly, we 
see that in this respect our model is in qualitative
agreement with effective QCD models. Note also the complementary discussion of
this point in~\cite{Aharony:2007uu}.

To conclude this section, we comment on the relation of the present results to
those of our previous paper \cite{Erdmenger:2007ap} where we considered a
constant non-Abelian gauge field background for zero quark mass. From equation
\eqref{eq:isospinIndices}, the difference between a constant non-vanishing
background gauge field and the varying one becomes clear. In
\cite{Erdmenger:2007ap} the field is chosen to be constant in~$\rho$ and terms
quadratic in the background gauge field~$\tilde A_0^3\ll 1$ are neglected. This 
implies that the square~$(\wn\mp\mn)^2$ in~\eqref{eq:eomX} and~\eqref{eq:eomY} is replaced by~$\wn^2\mp 2\wn\mn$, 
such that we obtain the indices~$\beta=\pm \wn
\sqrt{1\mp\frac{\bar A^3_0(\rho=1)}{(2\pi T) \wn}}$ instead of~\eqref{eq:isospinIndices}. If we
additionally assume $\wn\ll \tilde A^3_0$, then the $1$ under the square root
can be neglected.
In this case the spectral function develops a non-analytic structure coming from
the $\sqrt \omega$ factor in the index.

However in the case considered here, the background gauge field is a
non-constant function of $\rho$ which vanishes at the horizon. Therefore the
indices have the usual form $\beta = \pm i \omega$ from
(\ref{eq:isospinIndices}), and there is no non-analytic behavior of the spectral
functions, at least none originating from the indices.

It will also be interesting to consider isospin diffusion in the setup of the
present paper. However, in order to see non-Abelian effects in the diffusion
coefficient, we need to give the background gauge field a more general direction
in flavor space or a dependence on further space-time coordinates
besides~$\rho$. In that case, we will have a non-Abelian term in the background
field strength~$\tilde F_{\mu\nu}=\partial_\mu\tilde A^a_\nu-\partial_\nu \tilde
A^a_\mu +f^{abc} \tilde A^b_\mu \tilde A^c_\nu {\varrho_H}^2/(2\pi\alpha')$ 
in contrast to~$\partial_\rho
\tilde A_0^a$ considered here.

\section{Conclusion}
\label{sec:conclusion}

Two distinct setups were examined here at non-zero charge density in the black
hole phase. First, switching on a baryon chemical potential at non-zero baryon
density, we find that nearly stable vector mesons exist close to the transition
line to the Minkowski phase. Far from this line, at small quark masses, we
essentially recover the picture given in the case of vanishing chemical
potential~\cite{Myers:2007we}. Increasing the quark mass beyond a distinct
value, the plasma changes its behavior in order to asymptotically behave as it
would at zero temperature. In the spectral functions we computed, this
zero-temperature-like behavior is found in form of line-like resonances, exactly
reproducing the zero-temperature supersymmetric vector meson mass spectrum. A
turning point at $m=m^{\text{turn}}$, where $m=\bar M/T$, is observed: 
Below $m^{\text{turn}}$, the resonance peaks move to
lower frequencies as function of rising quark mass. This
behavior of the system resembles the behavior known of that system
without a chemical potential~\cite{Myers:2007we}. Above
$m^{\text{turn}}$, the resonance peaks move to higher frequencies 
as function of the quark
mass. This is the zero-temperature-like regime. Moreover, an examination of the
diffusion coefficient reveals that the phase transition separating two different
black hole phases~\cite{Kobayashi:2006sb} is shifted towards smaller temperature
as the baryon density is increased.

Second, we switched on a nonzero isospin density, and equivalently an isospin
chemical potential arises. The spectral functions in this case show a
qualitatively similar behavior as those for baryonic potential. However, we
additionally observe a splitting of the single resonance peak at vanishing
isospin potential into three distinct resonances. This suggests that by
explicitly breaking the flavor symmetry by a chemical potential, the isospin
triplet states, vector mesons in our case, show a mass splitting similar to that
observed for QCD~\cite{He:2005nk}. It is an interesting task to explore the
features of this isospin theory in greater detail in order to compare with
available lattice data and effective QCD
models~\cite{Kogut:2002zg,Kogut:2002tm,Kogut:2004zg,Splittorff:2000mm,Loewe:2002tw,
Barducci:2004tt,Sannino:2002wp,Ebert:2005cs,Ebert:2005wr}.
In most of these approaches, baryon and isospin chemical potential are
considered at the same time, which suggests another promising extension of this
work. Moreover, in the context of gravity duals, it will be interesting to
compare our results for the isospin chemical potential to the recent work
\cite{Aharony:2007uu}.

Alternatively, instead of giving the gauge field time component a non-vanishing
vev, one may also switch on B-field components and connect the framework
developed in~\cite{Erdmenger:2007bn,Albash:2007bk,Albash:2007bq} with the
calculation of spectral functions for the dual gauge theory.

\begin{acknowledgments}
\label{sec:ack}
We are grateful to P.~Kerner, C.~Greubel, K.~Landsteiner,
D.~Mateos, G.~Policastro, A.~Starinets, L.~Yaffe and M.~Zagermann for useful
discussions and correspondence, as well as to R.~Myers for suggesting to
consider a $\rho$-dependent $A_0$ component in the isospin case.

Part of this work was funded by the \emph{Cluster of Excellence for
Fundamental Physics -- Origin and Structure of the Universe}.
\end{acknowledgments}

\begin{appendix}

\section{Notation}
\label{sec:notation}
The five-dimensional $AdS$ Schwarzschild black hole space in which we
work is  endowed with a metric of
signature $(-,+,+,+,+)$, as  given explicitly in (\ref{eq:AdSBHmetric}).
We make use of the Einstein notation to indicate sums over Lorentz indices,
and additionally simply sum over non-Lorentz indices, such as
 gauge group indices, whenever they occur twice in a term.

To distinguish between vectors in different dimensions of the $AdS$ space, we
use bold symbols like $\bm{q}$ for vectors in the the \emph{three spatial
dimensions} which do not live along the radial $AdS$ coordinate.
\emph{Four-vectors} which do not have components along the radial $AdS$
coordinates are denoted by symbols with an arrow on top, as $\vec{q}$.

The Green functions $G=\langle J I  \rangle$ considered give correlations
between currents $J$ and $I$. These currents couple to fields $A$ and 
$B$ respectively. In our notation we use symbols such as $G_{A^a_k A^b_l}$ to
denote correlators of currents coupling to fields $A^a_k$ and $A^b_l$, with
flavor indices $a,b$ and Lorentz indices $k,l=0,1,2,3$. If no other indices are
of relevance for the discussion we restrict ourselves to Lorentz indices. For
the gauge field combinations $X$ and $Y$ given in
(\ref{eq:flavorTrafo}), we obtain Green functions $G_{XY}$ or $G_{YX}$ denoting
correlators of the corresponding currents.

\end{appendix}  
\providecommand{\href}[2]{#2}\begingroup\raggedright\endgroup


\end{document}